\documentclass[%
reprint, 
nofootinbib,
amsmath,amssymb,
aps, 
prl,
floatfix,
]{revtex4-2}

\usepackage[colorlinks=true,
citecolor=blue,
linkcolor=blue,
urlcolor=blue,
filecolor=blue,
pdfborder={0 0 0}]{hyperref}
\usepackage[capitalize]{cleveref}
\usepackage{physics}
\usepackage{soul}

\usepackage[dvipsnames]{xcolor}
\usepackage{tikz}
\usepackage{adjustbox}
\usetikzlibrary{decorations.markings,decorations.text,decorations.pathmorphing,decorations.pathreplacing,
	arrows.meta,positioning,intersections,calc,fit}
\usetikzlibrary{mindmap,backgrounds,trees,shapes.geometric}

\def\mZ{\mathcal{Z}}
\def\mO{\mathcal{O}}
\def\sX{\ensuremath{\overline{X}}}
\def\sA{\overline{A}}

\def\se{\ensuremath{S^{(2)}_A} }

\begin{document}

	\title{From Quantum Circuits with Ultraslow Dynamics to Classical Plaquette Models}
	
	\author{Vikram Ravindranath}
	\email{vikram.ravindranath@bc.edu}
	\author{Hanchen Liu}
	\author{Xiao Chen}
	\email{chenaad@bc.edu}
	\affiliation{Department of Physics, Boston College, Chestnut Hill, MA 02467, USA}
	
	\begin{abstract}
		We introduce a family of hybrid quantum circuits involving unitary gates and projective measurements that display a measurement-induced phase transition. Remarkably, the volume-law phase featuring logarithmic entanglement growth for certain initial states. We attribute this slow entanglement growth to the similarly slow growth of the participation entropy, which bounds the entanglement. Furthermore, the quantum circuit can be mapped to a classical spin model with real positive Boltzmann weights which involves local multi-spin interactions and displays glassy dynamics at finite temperature. We trace the origin of both the slow quantum dynamics and the classical glassiness to the presence of large, non-local symmetry operators. Our work establishes a novel connection between quantum entanglement dynamics and classical glassy behavior, offering a new geometric perspective on  entanglement phase transitions.
	\end{abstract}
	
	\maketitle
	
	In recent years, there has been significant progress in the study of monitored quantum dynamics, where unitary evolution is interspersed with repeated measurements. A key discovery is the emergence of measurement-induced entanglement phase transitions (MIPTs) \cite{MIPT0,MIPT1,MIPT2,MIPT3}. These transitions are well captured by hybrid quantum circuit models that combine unitary gates with measurement gates. At low measurement rates, the system exhibits volume-law entanglement: the entanglement entropy grows linearly in time and eventually saturates at a value proportional to system size. As the measurement rate increases, the system undergoes a transition to a disentangled area-law phase. The MIPT has also been interpreted through the lens of quantum error correction \cite{Choi_2020,Li_2021,Fan_2021,Gullans_2020}, motivating broader exploration of measurement-induced phenomena \cite{sang2021mphase,Lavasani_2021,Fisher_2023}.

	Motivated by these developments,  we introduce a class of hybrid random quantum circuits composed of both unitary and projective measurement gates. The unitary gates are restricted to those that map computational basis states to other basis states, up to a phase factor. Each measurement consists of a {\it pair} of single-qubit X and Z measurements acting on different sites.
	
	As expected, our model exhibits an entanglement phase transition between volume-law and area-law phases. However, it reveals a novel and intriguing feature: unlike previous studies where entanglement entropy quickly saturates to the volume-law value, our model displays logarithmically slow entanglement growth for certain initial states. This results in an exponentially long timescale to reach the steady-state entanglement entropy. We attribute this behavior to the ultraslow logarithmic growth of the participation entropy (PE), which serves as an upper bound for the entanglement entropy.
	
	Interestingly, we find that one example of 1+1D hybrid quantum dynamics can be mapped, via a transfer matrix approach, to a classical 2D random plaquette model with multi-spin interaction at zero temperature. This classical model—previously studied in Ref.~\cite{Liu24}—exhibits a phase transition in the structure of its symmetry operators as a function of system parameters. We establish a rigorous correspondence between this transition and the MIPT by using the replica method. In the volume-law phase, the associated symmetry operator is extensive, spreading over the entire system. Numerical simulations further confirm that this leads to a finite-temperature phase with glassy dynamics.
	
	Notably, this finite-temperature classical model can be mapped back to quantum dynamics with weak measurements, thereby establishing a striking connection between highly entangled volume-law phases, slow quantum dynamics, and glassy behavior in classical statistical models.
	
\begin{figure}
		\centering
		\includegraphics[width=\linewidth]{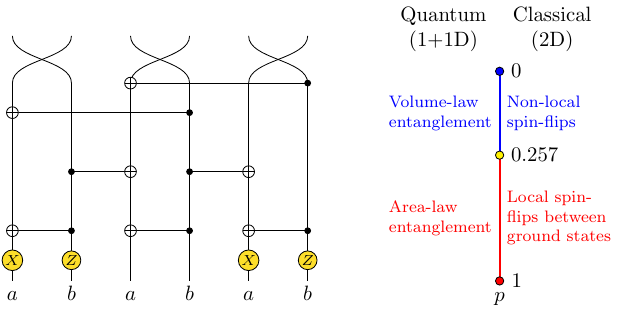}
		\caption{(Left) A single time-step of the circuit with 3 unit cells. Each cell consists of two qubits, labelled $a$ and $b$. First, both $a$ and $b$ in a given cell are measured (yellow circles) in the $X$ and $Z$ directions, at a rate $p$ per unit cell. CNOT gates are then applied such that for each qubit $b$ ($\bullet$) located in unit cell $j$, the targets ($\oplus$) are the $a$ qubits at $j-1, j$ and  $j+1$. Finally, $a$ and $b$ are swapped within each unit cell. (Right) A schematic of the entanglement phases in the quantum circuit and their corresponding phases in the 2D classical spin model.}
		\label{fig:circ_u}
	\end{figure}

	\textit{Model -- } The one-dimensional model consists of $L$ unit cells, each containing two qubits labelled $a$ and $b$, resulting in a total of $N_q = 2L$ qubits. These qubits undergo time evolution involving \textit{uniform} unitary gates and projective measurements. First, at a rate $p$ per time step, each unit cell undergoes projective measurements along the $X(Z)$ axes on the $a(b)$ qubit. CNOT  gates are then applied between $b$ and $a$ qubits within as well as across nearest neighbor unit cells. Finally, the $a$ and $b$ qubits in a unit cell are swapped, and the process repeats. One timestep of this circuit is depicted in \cref{fig:circ_u}.
	
	The circuit is a Clifford circuit which maps each string of Pauli operators to another string. For a state $\ket{\psi}$, we study the evolution of the $N_q$ independent Pauli strings $\qty{S_j}$ (the ``stabilizers") which obey $\comm{S_i}{S_j} = 0$ and $S_j\ket{\psi}=\ket{\psi}$. These stabilizers --- referred to collectively as a ``tableau" --  can be efficiently simulated \cite{Aaronson_2004}. Additionally, the circuit maps strings of $X$ ($Z$) operators only to other $X$ ($Z$) strings. We define \( N_X \) (\( N_Z \)) as the number of \( X \) (\( Z \)) stabilizers, with the constraint \( N_X + N_Z = N_q \). While the unitary gates preserve \( N_X \) and \( N_Z \), measurements can modify their values.

	\textit{Results --} We study the dynamics and steady-state behavior of the entanglement entropy $S_A = -\Tr(\rho_A\log_2\rho_A)$, where $\rho_A=\Tr_{\overline{A}}\rho$ is the reduced density matrix describing a subsystem $A$; $\rho$ is the density matrix of the entire system. The entanglement entropy of Clifford states depends on the stabilizers, but \textit{not} on their sign. Moreover, by the linearity of operator evolution, stabilizers that differ only by a sign map to operators that too only differ in sign. We will henceforth distinguish initial conditions only based on their stabilizers, e.g., the single-qubit states $\ket{0}$ and $\ket{1}$ are equivalent, but not $\ket{1}$ and $\ket{+}$.
	
	We begin by studying the evolution of product states which are eigenstates of $Z (X)$ on the $a (b)$ sublattice. As $p$ is increased, we find a transition from a volume-law phase $(S_A\sim |A|)$ with ballistic entanglement growth, to an area-law phase $(S_A\sim \rm{const})$. This behavior does not change even if the initial states are randomly chosen to be stabilized by $Z$ and $X$, provided that they exist at equal density. Further details of these results are presented in the Supplementary Material (SM, \cite{supp}).

	\textit{Slow Dynamics -- } Surprisingly, in the volume-law phase, we observe logarithmic entanglement growth for certain initial conditions. For instance, if one considers uniform initial stabilizers (all $Z$ or $X$), the entanglement growth in the volume-law phase is logarithmic $S_A(t) \sim\log t$ instead of ballistic\footnote{Such initial conditions lead to no entanglement growth in the absence of measurements, since CNOT gates do not entangle e.g.~$\ket{++}$ or $\ket{01}$.}. Moreover, slow dynamics is observed for \textit{any} random initial state  provided $|N_Z(t=0) - N_X(t=0)| \sim \mO\qty(L)$.  In these cases, the entanglement first grows linearly for a duration proportional to $N_q-|N_Z(t=0) - N_X(t=0)|$,  followed by a slow logarithmic growth. These ultraslow dynamics are in opposition to the widely held belief that hybrid circuits always scramble rapidly in the volume-law phase. The phenomenology of the area-law phase meanwhile remain unchanged.

	\begin{figure}
		\centering
		\includegraphics[width=\linewidth]{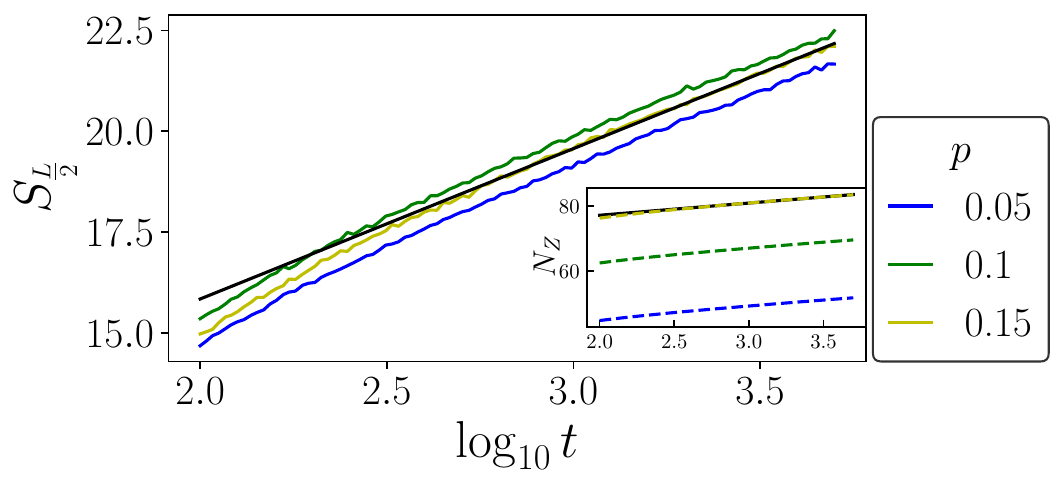}
		\caption{The time evolution of the half-chain entanglement entropy $S_{\frac{L}{2}}$, shown on a logarithmic scale, beginning from a uniform initial state of $L=100$ sites (the stabilizers are all single-site $X$), for various $p<p_c$. The entanglement growth is logarithmic in time, and \textit{non-monotonic} with increasing $p$. A fit yields a universal slope $S_{\frac{L}{2}}\sim 3.7\log_{10} t + c(p)$. (Inset) The participation entropy, in this case given by $N_Z$, also grows logarithmically, with the \textit{same slope}, as predicted.}
		\label{fig:slow_dyn}
	\end{figure}

	The slow dynamics can be understood as follows. The entanglement entropy \( S_A \) of a quantum state is bounded by its PE, which characterizes how ``spread out'' the state is in a given basis. Specifically, if $\ket{\psi}=\sum\limits_{s\in S} C_s \ket{s}$ in some single-qubit product basis $B$, where $S$ labels a subset of $B$, then
	\begin{equation}
		PE_B(\ket{\psi}) = \log |S|.
		\label{eq:PE}
	\end{equation}
	For the wavefunctions of interest here, $PE_{X/Z}$ in the \( X \) (\( Z \)) basis is simply given by \( N_Z \) (\( N_X \)), and therefore,
	\begin{equation}
		S_A \leq \min\qty(N_X, N_Z).
		\label{eq:up_bound}
	\end{equation}
	
	Consider an initial state with all qubits polarized in the \( \pm X \) direction. In this case, \( S_A \) is bounded by \( N_Z \). At \( t = 0 \), we have \( S_A = N_Z = 0 \). However, the number of \( Z \) stabilizers can rapidly increase due to \( Z \) measurements. In the volume-law phase, after applying a finite layer of measurements, we have \( N_Z = \alpha L \) with \( \alpha \ll 1 \). These \( Z \) stabilizers, which initially correspond to single-site Pauli \( Z \) operators, grow ballistically under the action of CNOT gates, eventually forming extensive \( Z \)-strings. 
	This growth also leads to a rapid increase of entanglement entropy up to \( \alpha L \) within the accessible Hilbert space.

	The presence of these extensive \( Z \) stabilizers suppresses further growth in \( N_Z \), leading to the prolonged logarithmic growth observed in the entanglement dynamics. This is because, in our model, each measurement step involves simultaneously measuring a pair of Pauli \( X \) and \( Z \) operators. At late times (\( t \sim O(L) \)), both \( X \) and \( Z \) stabilizers typically have extensive spatial support. In such cases, a single-qubit \( X \)/\( Z \) measurement will generally not commute with at least one of the existing \( Z \)/\( X \) stabilizers. As a result, applying an additional \( X \)/\( Z \) measurement pair typically does not increase \( N_Z \).
	
	To increase \( N_Z \) by one, the \( X \) measurement must occur on a site that is not part of any existing \( Z \) stabilizer. The probability of this happening is approximately:
	\begin{align}
		P = \qty(1 - \frac{l}{L})^{N_Z},
		\label{eq:PNX}
	\end{align}
	where \( l \) is the typical length of a stabilizer.
	
	This probability decays exponentially with \( N_Z \), implying that the typical time (inverse of $P$) required to increase \( N_Z \) becomes exponentially long. In particular, if we assume that stabilizers on average occupy $l=3L/4$ sites\footnote{This is a reasonable assumption given that the unitary dynamics consists of CNOT gates which lead to $XX\leftrightarrow X$ and $ZZ\leftrightarrow Z$.}, we expect the coefficient in front of $\log_{10} t$ is $-\log(10)\log(1-l/L)\approx 3.19$ which is a universal constant independent of measurement rate and the initial state. While the observed coefficient is slightly larger (3.7), it is  independent of both the measurement rate and the initial state (Fig.~\ref{fig:slow_dyn}). This explains the observed logarithmic growth of both \( N_Z \) and the entanglement entropy \( S_A \). A similar phenomenon has been observed in the exponentially long purification time characteristic of generic volume-law phases~\cite{Gullans_2020}.

	Based on the above analysis, we can construct a generic random quantum circuit that exhibits slow dynamics when initialized in a special state with low participation entropy. There should be a basis in which PE remains invariant under unitary evolution, so each basis state is mapped to another basis state (up to a phase). Pairs of qubits are then measured in non-commuting (e.g.~$X$ and $Z$) bases. We numerically confirm this behavior in certain hybrid random Clifford circuits, with results presented in the SM\cite{supp}.

	\textit{Connections to a 2d classical model -- }
	Returning to the model we defined in Fig.~\ref{fig:circ_u}, we write the wavefunction $\ket{\psi}$ emerging from the circuit at time $T$ as a superposition over Z-basis states
	\begin{equation}
		\ket{\psi\qty(T)} \propto \sum\limits_{s_T}\mZ\qty(s_T)\ket{s_T}.
		\label{eq:wavefn}
	\end{equation}
	Each basis state is labeled by a string of $2L$ bits, denoted by $s_T$, and $\mZ(s_T)$ is proportional to the corresponding amplitude in $\ket{\psi}$. The bitstring $s_T$ is further divided into two sublattices $a$, $b$ with each bit labeled as $s^a_{x,T}$ and $s^b_{x,T}$ respectively. In our circuits, $\mZ$ is \textit{real and nonnegative} for each $s_T$, so we can interpret $\mZ$ as the partition function of a classical system, governed by a circuit-realization-dependent Hamiltonian $H_c$\footnote{In the discussion the follows, we consider the degrees of freedom at each site to be a bit $s\in\qty{0,1}$ as opposed to conventional Ising variables $\sigma=\pm1$, but the physics is entirely that of Ising spin systems, with the identification $\sigma=(-1)^s$.}.

    \begin{figure}
		\centering
		\includegraphics[width=\linewidth]{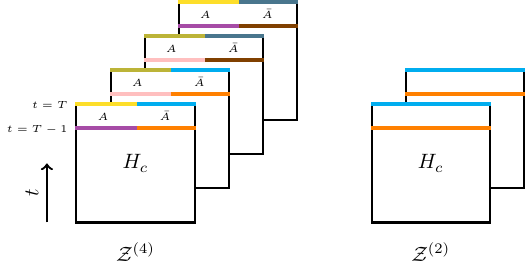}
		\caption{The boundary conditions that are imposed at rows $T-1$ and $T$ across multiple copies of $H_c(\qty{q})$, to calculate $\mathcal{Z}^{(4/2)}$. Spins in regions of the same colour are constrained to be equal. $\mathcal{Z}^{(4)}$ requires different constraints for $A$ and $\bar{A}$, while they are identical for $\mathcal{Z}^{(2)}$.}
		\label{fig:BCs}
	\end{figure}

    \begin{figure*}[t]
		\centering
		\includegraphics{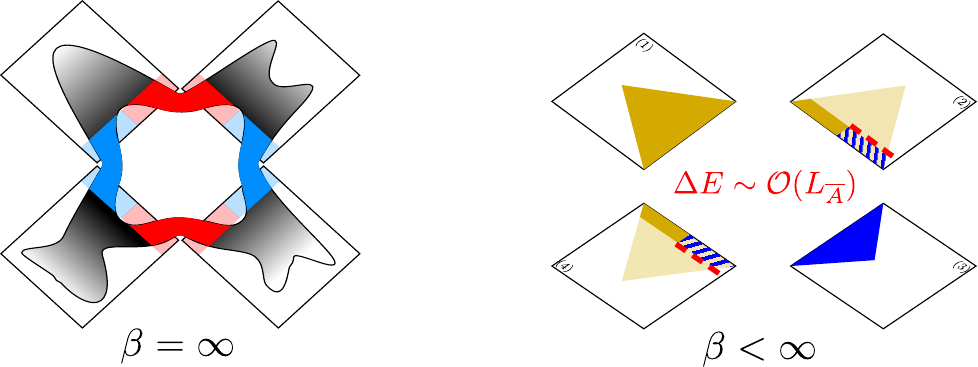}
		\caption{The ground state configurations (left) and the boundary excitations (right) which contribute to $\se$. When $\beta=\infty$, the contributions come from $\sX$ that span subsystems $A$ (red) \textbf{and} $\sA$ (blue), and are identical across subsystems that are connected by shaded lines. When $\beta<\infty$, boundary excitations begin to contribute. This involves choosing $\sX$s independently on (1) (blue) and (3) (gold). An $\mO(L)$ energy cost (red, dashed lines) is incurred from the mismatched boundaries (blue shading) on (2) and (4).}
		\label{fig:sym-op}
	\end{figure*}

	$H_c$ describes a system of boolean variables or bits $s_{(x,t)}\in\qty{0,1}$ situated on the vertices of a $L\times T$ rectangular lattice. Each term in $H_c$ is a modulo 2 sum of $q$ bits, with $q_{x,t} \in \qty{1,5}$ being vertex dependent.
	\begin{equation}
		\begin{aligned}
			\includegraphics[width=\linewidth]{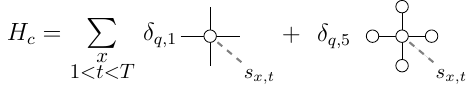}.
		\end{aligned}
	\end{equation}
	At every vertex $(x,t)$ $(1<t<T)$, the sum involves the $q_{x,t}$ bits marked $\circ$\footnote{In terms of spin variables, this corresponds to either a magnetic field at each point, or an Ising coupling involving the 5 spins marked $\circ$.}. The bits at rows $T-1$ and $T$ are fixed to be $s_{x,T-1} = s^b_{x,T}$ and $s_{x,T}= s^a_{x,T}$, while the boundary conditions on the first two rows are determined by the initial quantum state. We will refer to the terms of $H_c$ as ``parity checks".
	
	The ground states of $H_c$ are those that satisfy every constraint, and the all-zero state with $s_{x,t}=0$ is always a trivial ground state in the absence of imposing boundary conditions. The nontrivial ground-states can be labelled by the locations of their non-zero bits. Each ground-state $g$ defines a product of bit-flip operators $\sX_g = \prod\limits_{(x,t)\in g} X_{x,t}$, with an $X$ operator on $(x,t)$ if $s_{x,t}=1$, and identity otherwise. The $\sX_g$ commute with the Hamiltonian $H_c$, thereby generating its symmetry group $G$\footnote{Even though we are discussing purely classical bits, we use the language of Pauli operators to simplify the discussion.}. At zero temperature, the partition function $\mZ=\lim\limits_{\beta\to\infty} e^{-\beta H_c}$ simply counts the number of ground states, so $\mZ=|G|$. One can choose $k\equiv\log_2|G|$ independent $\sX_j$ as a basis to generate $|G|$ and thus, all the ground states of $H_c$.

	This model undergoes a structural phase transition in the behavior of $\sX_j$ as $p$ is tuned \cite{Liu24}. For $p<p_c$, $\sX_j$  is a two-dimensional operator that flips an extensive number of spins. However, when the single-body term has a rate $p>p_c$, $\sX_j$ can be made local and can only flip spins over finite number of sites.
	
	To further establish the connection between this transition and the MIPT, we consider the second R\'{e}nyi entropy $S^{(2)}_A \equiv -\log\Tr(\rho^2_A)$ which can be written as
	\begin{equation}
		S^{(2)}_A = -\log \mZ^{(4)} + 2\log \mZ^{(2)}.
	\end{equation}
	$\mZ^{(4/2)}$ is the partition function of 4 or 2 copies of one realization of the system, identically governed by $H_c(\qty{q})$ in the bulk. The boundary conditions on the top rows, described graphically in \cref{fig:BCs}, enforce equality between the bits on either the $A$ or $\overline{A}$ subregions across replicas. The number of ground states is equal to the number of solutions to all the constraints (both from $H_c$ and the boundaries), so $\log \mZ^{(4/2)}\equiv k^{(4/2)}$ is the number of independent $\sX$. Thus, $S^{(2)}_A = 2k^{(2)} - k^{(4)}$ is the difference in the number of independent $\sX$, under different boundary conditions. 
	
\begin{figure}
			\centering
			\includegraphics[width=\linewidth]{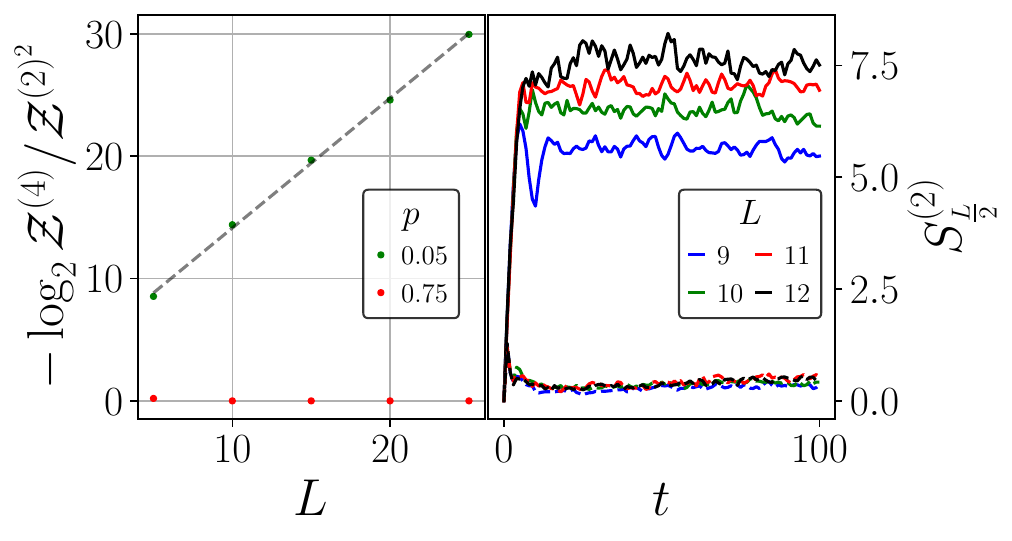}
			\caption{Insights from the classical mapping. (Left) Scaling of entanglement entropy in the volume-law (green) and area-law (red) phases, as obtained from $H_c$ at $\beta=\infty$. $L$ is the linear size of the system. (Right) $S^{(2)}_\frac{L}{2} (t)$ at fixed $p=0.1$ for $\beta=2.0$ (solid lines) and $\beta=0.5$ (dashed lines), describing the volume- and area-law phases, respectively.}
			\label{fig:stat_map}
		\end{figure}

	The independent \(\sX\)  can be classified into two types: bulk operators $G_B$, which reside entirely within the bulk and do not touch the top boundary, and boundary operators which terminate nontrivially at the top boundary. The quantities \( \log \mathcal{Z}^{(4)} \) and \( \log \mathcal{Z}^{(2)} \) share the same bulk $\sX$ but differ in their boundary operators. In \( \log \mathcal{Z}^{(2)} \), two boundary operators that share the same boundary configuration are glued together and therefore, treated as a symmetry operator of $\mZ^{(2)}$. In contrast, \( \log \mathcal{Z}^{(4)} \) requires more complex boundary conditions gluing $A$ and $\overline{A}$ across copies; ultimately, only the $\sX$ which can be localized exclusively outside $A$ or $\overline{A}$, contribute. Thus, \( S^{(2)}_A \) is equal to the number of independent boundary operators that span both \( A \) and \( \sA \). \cref{fig:sym-op} shows the operators that contribute to $S^{(2)}_A$. When \( p < p_c \), most boundary $\sX$s span both $A$ and $\overline{A}$. Moreover, the \textit{number} of these operators scales with the length of the boundary, resulting in $S^{(2)}_A\sim |A|$ (volume-law). On the other hand, when \( p > p_c \), nearly all boundary operators can be localized on either \( A \) and \( \bar{A} \), resulting in an area-law scaling. This behavior is confirmed numerically in \cref{fig:stat_map} (left).

	\textit{Weak measurement -- }
The classical model at finite temperature can be mapped back to a quantum circuit with weak forced measurements through the following modification:

	\begin{enumerate}
		\item Projective Z-measurements are replaced by a forced Z-measurement of the form $e^{\beta Z}$ (the X-measurements are unchanged),
		\item \textit{Every} $a$ site additionally undergoes a forced X-measurement of the form $e^{\gamma X}$, where $e^{-2\gamma} \equiv \tanh(\beta)$.
	\end{enumerate}

	Since this is no longer a Clifford circuit, we simulate the dynamics using MPS techniques. As shown in Fig.~\ref{fig:stat_map}, when $p<p_c$, the system at large $\beta$ exhibits a phase with volume-law entanglement scaling. In particular, the slow logarithmic growth of entanglement persists for certain initial states. However, as $\beta$ decreases, we observe a phase transition to an area-law phase. This transition can be interpreted as a finite-temperature phase transition of the corresponding classical model in replica space. Specifically, the boundary symmetry operator in each copy can produce a mismatch near the boundary with an energy cost of order $O(L)$ (See \cref{fig:sym-op}). Because there are $O(\exp(L))$ such boundary excitations, a free-energy argument shows that they can drive a finite-temperature phase transition. A detailed explanation is provided in the SM \cite{supp}.

		\textit{Conclusion -- }
		We study a family of 1+1D hybrid quantum circuits that exhibit MIPT and demonstrate that, in the volume-law phase, the entanglement entropy can grow logarithmically with time when evolved from certain special initial states. We attribute this behavior to the similarly slow, logarithmic growth of the participation entropy. Notably, such slow dynamics has been observed in certain strongly fragmented systems with boundary perturbations. 
		
		Interestingly, we find that one sign-free non-unitary quantum dynamics studied in this work can be mapped to a two-dimensional classical model with multi-spin interactions. In this mapping, the MIPT corresponds to a structural phase transition of the symmetry operator in the classical model. In the phase where the symmetry operator is extensive, spin flips between distinct ground states are strongly suppressed—resulting in glassy dynamics in the classical system at finite temperature.
		
		While the transfer matrix approach has previously been used to relate quantum systems to higher-dimensional classical models, our work highlights a novel perspective: it demonstrates how highly entangled quantum states and slow quantum dynamics can be understood through classical models exhibiting glassy behavior. Looking ahead, we plan to explore the finite-temperature (weak measurement) regime in greater detail and examine its potential implications for both quantum and classical error-correcting codes.
		
		\textit{ Acknowledgments-- }
		We gratefully acknowledge the computing resources provided by Research Services at Boston College. This research was supported in part by the National Science Foundation under Grant No. DMR-2219735.

		\nocite{ZX1,ZX2,ZX3,Hong2025}
        \bibliography{reference}

\begin{thebibliography}{18}%
\makeatletter
\providecommand \@ifxundefined [1]{%
 \@ifx{#1\undefined}
}%
\providecommand \@ifnum [1]{%
 \ifnum #1\expandafter \@firstoftwo
 \else \expandafter \@secondoftwo
 \fi
}%
\providecommand \@ifx [1]{%
 \ifx #1\expandafter \@firstoftwo
 \else \expandafter \@secondoftwo
 \fi
}%
\providecommand \natexlab [1]{#1}%
\providecommand \enquote  [1]{``#1''}%
\providecommand \bibnamefont  [1]{#1}%
\providecommand \bibfnamefont [1]{#1}%
\providecommand \citenamefont [1]{#1}%
\providecommand \href@noop [0]{\@secondoftwo}%
\providecommand \href [0]{\begingroup \@sanitize@url \@href}%
\providecommand \@href[1]{\@@startlink{#1}\@@href}%
\providecommand \@@href[1]{\endgroup#1\@@endlink}%
\providecommand \@sanitize@url [0]{\catcode `\\12\catcode `\$12\catcode
  `\&12\catcode `\#12\catcode `\^12\catcode `\_12\catcode `\%12\relax}%
\providecommand \@@startlink[1]{}%
\providecommand \@@endlink[0]{}%
\providecommand \url  [0]{\begingroup\@sanitize@url \@url }%
\providecommand \@url [1]{\endgroup\@href {#1}{\urlprefix }}%
\providecommand \urlprefix  [0]{URL }%
\providecommand \Eprint [0]{\href }%
\providecommand \doibase [0]{https://doi.org/}%
\providecommand \selectlanguage [0]{\@gobble}%
\providecommand \bibinfo  [0]{\@secondoftwo}%
\providecommand \bibfield  [0]{\@secondoftwo}%
\providecommand \translation [1]{[#1]}%
\providecommand \BibitemOpen [0]{}%
\providecommand \bibitemStop [0]{}%
\providecommand \bibitemNoStop [0]{.\EOS\space}%
\providecommand \EOS [0]{\spacefactor3000\relax}%
\providecommand \BibitemShut  [1]{\csname bibitem#1\endcsname}%
\let\auto@bib@innerbib\@empty
\bibitem [{\citenamefont {Li}\ \emph {et~al.}(2018)\citenamefont {Li},
  \citenamefont {Chen},\ and\ \citenamefont {Fisher}}]{MIPT0}%
  \BibitemOpen
  \bibfield  {author} {\bibinfo {author} {\bibfnamefont {Y.}~\bibnamefont
  {Li}}, \bibinfo {author} {\bibfnamefont {X.}~\bibnamefont {Chen}},\ and\
  \bibinfo {author} {\bibfnamefont {M.~P.~A.}\ \bibnamefont {Fisher}},\
  }\bibfield  {title} {\bibinfo {title} {Quantum zeno effect and the many-body
  entanglement transition},\ }\href
  {https://doi.org/10.1103/PhysRevB.98.205136} {\bibfield  {journal} {\bibinfo
  {journal} {Phys. Rev. B}\ }\textbf {\bibinfo {volume} {98}},\ \bibinfo
  {pages} {205136} (\bibinfo {year} {2018})}\BibitemShut {NoStop}%
\bibitem [{\citenamefont {Li}\ \emph {et~al.}(2019)\citenamefont {Li},
  \citenamefont {Chen},\ and\ \citenamefont {Fisher}}]{MIPT1}%
  \BibitemOpen
  \bibfield  {author} {\bibinfo {author} {\bibfnamefont {Y.}~\bibnamefont
  {Li}}, \bibinfo {author} {\bibfnamefont {X.}~\bibnamefont {Chen}},\ and\
  \bibinfo {author} {\bibfnamefont {M.~P.~A.}\ \bibnamefont {Fisher}},\
  }\bibfield  {title} {\bibinfo {title} {Measurement-driven entanglement
  transition in hybrid quantum circuits},\ }\href
  {https://doi.org/10.1103/PhysRevB.100.134306} {\bibfield  {journal} {\bibinfo
   {journal} {Phys. Rev. B}\ }\textbf {\bibinfo {volume} {100}},\ \bibinfo
  {pages} {134306} (\bibinfo {year} {2019})}\BibitemShut {NoStop}%
\bibitem [{\citenamefont {Skinner}\ \emph {et~al.}(2019)\citenamefont
  {Skinner}, \citenamefont {Ruhman},\ and\ \citenamefont {Nahum}}]{MIPT2}%
  \BibitemOpen
  \bibfield  {author} {\bibinfo {author} {\bibfnamefont {B.}~\bibnamefont
  {Skinner}}, \bibinfo {author} {\bibfnamefont {J.}~\bibnamefont {Ruhman}},\
  and\ \bibinfo {author} {\bibfnamefont {A.}~\bibnamefont {Nahum}},\ }\bibfield
   {title} {\bibinfo {title} {Measurement-induced phase transitions in the
  dynamics of entanglement},\ }\href
  {https://doi.org/10.1103/PhysRevX.9.031009} {\bibfield  {journal} {\bibinfo
  {journal} {Phys. Rev. X}\ }\textbf {\bibinfo {volume} {9}},\ \bibinfo {pages}
  {031009} (\bibinfo {year} {2019})}\BibitemShut {NoStop}%
\bibitem [{\citenamefont {Chan}\ \emph {et~al.}(2019)\citenamefont {Chan},
  \citenamefont {Nandkishore}, \citenamefont {Pretko},\ and\ \citenamefont
  {Smith}}]{MIPT3}%
  \BibitemOpen
  \bibfield  {author} {\bibinfo {author} {\bibfnamefont {A.}~\bibnamefont
  {Chan}}, \bibinfo {author} {\bibfnamefont {R.~M.}\ \bibnamefont
  {Nandkishore}}, \bibinfo {author} {\bibfnamefont {M.}~\bibnamefont
  {Pretko}},\ and\ \bibinfo {author} {\bibfnamefont {G.}~\bibnamefont
  {Smith}},\ }\bibfield  {title} {\bibinfo {title} {Unitary-projective
  entanglement dynamics},\ }\href {https://doi.org/10.1103/PhysRevB.99.224307}
  {\bibfield  {journal} {\bibinfo  {journal} {Phys. Rev. B}\ }\textbf {\bibinfo
  {volume} {99}},\ \bibinfo {pages} {224307} (\bibinfo {year}
  {2019})}\BibitemShut {NoStop}%
\bibitem [{\citenamefont {Choi}\ \emph {et~al.}(2020)\citenamefont {Choi},
  \citenamefont {Bao}, \citenamefont {Qi},\ and\ \citenamefont
  {Altman}}]{Choi_2020}%
  \BibitemOpen
  \bibfield  {author} {\bibinfo {author} {\bibfnamefont {S.}~\bibnamefont
  {Choi}}, \bibinfo {author} {\bibfnamefont {Y.}~\bibnamefont {Bao}}, \bibinfo
  {author} {\bibfnamefont {X.-L.}\ \bibnamefont {Qi}},\ and\ \bibinfo {author}
  {\bibfnamefont {E.}~\bibnamefont {Altman}},\ }\bibfield  {title} {\bibinfo
  {title} {Quantum error correction in scrambling dynamics and
  measurement-induced phase transition},\ }\href
  {https://doi.org/10.1103/PhysRevLett.125.030505} {\bibfield  {journal}
  {\bibinfo  {journal} {Phys. Rev. Lett.}\ }\textbf {\bibinfo {volume} {125}},\
  \bibinfo {pages} {030505} (\bibinfo {year} {2020})}\BibitemShut {NoStop}%
\bibitem [{\citenamefont {Li}\ and\ \citenamefont {Fisher}(2021)}]{Li_2021}%
  \BibitemOpen
  \bibfield  {author} {\bibinfo {author} {\bibfnamefont {Y.}~\bibnamefont
  {Li}}\ and\ \bibinfo {author} {\bibfnamefont {M.~P.~A.}\ \bibnamefont
  {Fisher}},\ }\bibfield  {title} {\bibinfo {title} {Statistical mechanics of
  quantum error correcting codes},\ }\href
  {https://doi.org/10.1103/PhysRevB.103.104306} {\bibfield  {journal} {\bibinfo
   {journal} {Phys. Rev. B}\ }\textbf {\bibinfo {volume} {103}},\ \bibinfo
  {pages} {104306} (\bibinfo {year} {2021})}\BibitemShut {NoStop}%
\bibitem [{\citenamefont {Fan}\ \emph {et~al.}(2021)\citenamefont {Fan},
  \citenamefont {Vijay}, \citenamefont {Vishwanath},\ and\ \citenamefont
  {You}}]{Fan_2021}%
  \BibitemOpen
  \bibfield  {author} {\bibinfo {author} {\bibfnamefont {R.}~\bibnamefont
  {Fan}}, \bibinfo {author} {\bibfnamefont {S.}~\bibnamefont {Vijay}}, \bibinfo
  {author} {\bibfnamefont {A.}~\bibnamefont {Vishwanath}},\ and\ \bibinfo
  {author} {\bibfnamefont {Y.-Z.}\ \bibnamefont {You}},\ }\bibfield  {title}
  {\bibinfo {title} {Self-organized error correction in random unitary circuits
  with measurement},\ }\href {https://doi.org/10.1103/PhysRevB.103.174309}
  {\bibfield  {journal} {\bibinfo  {journal} {Phys. Rev. B}\ }\textbf {\bibinfo
  {volume} {103}},\ \bibinfo {pages} {174309} (\bibinfo {year}
  {2021})}\BibitemShut {NoStop}%
\bibitem [{\citenamefont {Gullans}\ and\ \citenamefont
  {Huse}(2020)}]{Gullans_2020}%
  \BibitemOpen
  \bibfield  {author} {\bibinfo {author} {\bibfnamefont {M.~J.}\ \bibnamefont
  {Gullans}}\ and\ \bibinfo {author} {\bibfnamefont {D.~A.}\ \bibnamefont
  {Huse}},\ }\bibfield  {title} {\bibinfo {title} {Dynamical purification phase
  transition induced by quantum measurements},\ }\href
  {https://doi.org/10.1103/PhysRevX.10.041020} {\bibfield  {journal} {\bibinfo
  {journal} {Phys. Rev. X}\ }\textbf {\bibinfo {volume} {10}},\ \bibinfo
  {pages} {041020} (\bibinfo {year} {2020})}\BibitemShut {NoStop}%
\bibitem [{\citenamefont {Sang}\ and\ \citenamefont
  {Hsieh}(2021)}]{sang2021mphase}%
  \BibitemOpen
  \bibfield  {author} {\bibinfo {author} {\bibfnamefont {S.}~\bibnamefont
  {Sang}}\ and\ \bibinfo {author} {\bibfnamefont {T.~H.}\ \bibnamefont
  {Hsieh}},\ }\bibfield  {title} {\bibinfo {title} {Measurement-protected
  quantum phases},\ }\href {https://doi.org/10.1103/PhysRevResearch.3.023200}
  {\bibfield  {journal} {\bibinfo  {journal} {Phys. Rev. Res.}\ }\textbf
  {\bibinfo {volume} {3}},\ \bibinfo {pages} {023200} (\bibinfo {year}
  {2021})}\BibitemShut {NoStop}%
\bibitem [{\citenamefont {Lavasani}\ \emph {et~al.}(2021)\citenamefont
  {Lavasani}, \citenamefont {Alavirad},\ and\ \citenamefont
  {Barkeshli}}]{Lavasani_2021}%
  \BibitemOpen
  \bibfield  {author} {\bibinfo {author} {\bibfnamefont {A.}~\bibnamefont
  {Lavasani}}, \bibinfo {author} {\bibfnamefont {Y.}~\bibnamefont {Alavirad}},\
  and\ \bibinfo {author} {\bibfnamefont {M.}~\bibnamefont {Barkeshli}},\
  }\bibfield  {title} {\bibinfo {title} {Measurement-induced topological
  entanglement transitions in symmetric random quantum circuits},\ }\href
  {https://doi.org/10.1038/s41567-020-01112-z} {\bibfield  {journal} {\bibinfo
  {journal} {Nature Physics}\ }\textbf {\bibinfo {volume} {17}},\ \bibinfo
  {pages} {342–347} (\bibinfo {year} {2021})}\BibitemShut {NoStop}%
\bibitem [{\citenamefont {Fisher}\ \emph {et~al.}(2023)\citenamefont {Fisher},
  \citenamefont {Khemani}, \citenamefont {Nahum},\ and\ \citenamefont
  {Vijay}}]{Fisher_2023}%
  \BibitemOpen
  \bibfield  {author} {\bibinfo {author} {\bibfnamefont {M.~P.}\ \bibnamefont
  {Fisher}}, \bibinfo {author} {\bibfnamefont {V.}~\bibnamefont {Khemani}},
  \bibinfo {author} {\bibfnamefont {A.}~\bibnamefont {Nahum}},\ and\ \bibinfo
  {author} {\bibfnamefont {S.}~\bibnamefont {Vijay}},\ }\bibfield  {title}
  {\bibinfo {title} {Random quantum circuits},\ }\href
  {https://doi.org/10.1146/annurev-conmatphys-031720-030658} {\bibfield
  {journal} {\bibinfo  {journal} {Annual Review of Condensed Matter Physics}\
  }\textbf {\bibinfo {volume} {14}},\ \bibinfo {pages} {335–379} (\bibinfo
  {year} {2023})}\BibitemShut {NoStop}%
\bibitem [{\citenamefont {Liu}\ and\ \citenamefont {Chen}(2024)}]{Liu24}%
  \BibitemOpen
  \bibfield  {author} {\bibinfo {author} {\bibfnamefont {H.}~\bibnamefont
  {Liu}}\ and\ \bibinfo {author} {\bibfnamefont {X.}~\bibnamefont {Chen}},\
  }\bibfield  {title} {\bibinfo {title} {Plaquette models, cellular automata,
  and measurement-induced criticality},\ }\href
  {https://doi.org/10.1088/1751-8121/ad82bc} {\bibfield  {journal} {\bibinfo
  {journal} {Journal of Physics A: Mathematical and Theoretical}\ }\textbf
  {\bibinfo {volume} {57}},\ \bibinfo {pages} {435003} (\bibinfo {year}
  {2024})}\BibitemShut {NoStop}%
\bibitem [{\citenamefont {Aaronson}\ and\ \citenamefont
  {Gottesman}(2004)}]{Aaronson_2004}%
  \BibitemOpen
  \bibfield  {author} {\bibinfo {author} {\bibfnamefont {S.}~\bibnamefont
  {Aaronson}}\ and\ \bibinfo {author} {\bibfnamefont {D.}~\bibnamefont
  {Gottesman}},\ }\bibfield  {title} {\bibinfo {title} {Improved simulation of
  stabilizer circuits},\ }\href {https://doi.org/10.1103/PhysRevA.70.052328}
  {\bibfield  {journal} {\bibinfo  {journal} {Phys. Rev. A}\ }\textbf {\bibinfo
  {volume} {70}},\ \bibinfo {pages} {052328} (\bibinfo {year}
  {2004})}\BibitemShut {NoStop}%
\bibitem [{sup()}]{supp}%
  \BibitemOpen
  \href@noop {} {}\bibinfo {note} {See Supplemental Material which includes
  Refs.~[15-18], for details on the connection to the classical model at finite
  temperature, additional data on more generic circuits and Monte-carlo
  simulations of glassy behaviour.}\BibitemShut {Stop}%
\bibitem [{\citenamefont {Coecke}\ and\ \citenamefont {Duncan}(2011)}]{ZX1}%
  \BibitemOpen
  \bibfield  {author} {\bibinfo {author} {\bibfnamefont {B.}~\bibnamefont
  {Coecke}}\ and\ \bibinfo {author} {\bibfnamefont {R.}~\bibnamefont
  {Duncan}},\ }\bibfield  {title} {\bibinfo {title} {Interacting quantum
  observables: categorical algebra and diagrammatics},\ }\href
  {https://doi.org/10.1088/1367-2630/13/4/043016} {\bibfield  {journal}
  {\bibinfo  {journal} {New Journal of Physics}\ }\textbf {\bibinfo {volume}
  {13}},\ \bibinfo {pages} {043016} (\bibinfo {year} {2011})}\BibitemShut
  {NoStop}%
\bibitem [{\citenamefont {Coecke}\ and\ \citenamefont {Kissinger}(2018)}]{ZX2}%
  \BibitemOpen
  \bibfield  {author} {\bibinfo {author} {\bibfnamefont {B.}~\bibnamefont
  {Coecke}}\ and\ \bibinfo {author} {\bibfnamefont {A.}~\bibnamefont
  {Kissinger}},\ }\bibfield  {title} {\bibinfo {title} {Picturing quantum
  processes},\ }in\ \href@noop {} {\emph {\bibinfo {booktitle} {Diagrammatic
  Representation and Inference}}},\ \bibinfo {editor} {edited by\ \bibinfo
  {editor} {\bibfnamefont {P.}~\bibnamefont {Chapman}}, \bibinfo {editor}
  {\bibfnamefont {G.}~\bibnamefont {Stapleton}}, \bibinfo {editor}
  {\bibfnamefont {A.}~\bibnamefont {Moktefi}}, \bibinfo {editor} {\bibfnamefont
  {S.}~\bibnamefont {Perez-Kriz}},\ and\ \bibinfo {editor} {\bibfnamefont
  {F.}~\bibnamefont {Bellucci}}}\ (\bibinfo  {publisher} {Springer
  International Publishing},\ \bibinfo {address} {Cham},\ \bibinfo {year}
  {2018})\ pp.\ \bibinfo {pages} {28--31}\BibitemShut {NoStop}%
\bibitem [{\citenamefont {van~de Wetering}(2020)}]{ZX3}%
  \BibitemOpen
  \bibfield  {author} {\bibinfo {author} {\bibfnamefont {J.}~\bibnamefont
  {van~de Wetering}},\ }\bibfield  {title} {\bibinfo {title} {Zx-calculus for
  the working quantum computer scientist},\ }\href@noop {} {\bibfield
  {journal} {\bibinfo  {journal} {arXiv preprint arXiv:2012.13966}\ } (\bibinfo
  {year} {2020})}\BibitemShut {NoStop}%
\bibitem [{\citenamefont {Hong}\ \emph {et~al.}(2025)\citenamefont {Hong},
  \citenamefont {Guo},\ and\ \citenamefont {Lucas}}]{Hong2025}%
  \BibitemOpen
  \bibfield  {author} {\bibinfo {author} {\bibfnamefont {Y.}~\bibnamefont
  {Hong}}, \bibinfo {author} {\bibfnamefont {J.}~\bibnamefont {Guo}},\ and\
  \bibinfo {author} {\bibfnamefont {A.}~\bibnamefont {Lucas}},\ }\bibfield
  {title} {\bibinfo {title} {Quantum memory at nonzero temperature in a
  thermodynamically trivial system},\ }\href
  {https://doi.org/10.1038/s41467-024-55570-7} {\bibfield  {journal} {\bibinfo
  {journal} {Nature Communications}\ }\textbf {\bibinfo {volume} {16}},\
  \bibinfo {pages} {316} (\bibinfo {year} {2025})}\BibitemShut {NoStop}%
\end{thebibliography}%


\begin{thebibliography}{13}%
\makeatletter
\providecommand \@ifxundefined [1]{%
 \@ifx{#1\undefined}
}%
\providecommand \@ifnum [1]{%
 \ifnum #1\expandafter \@firstoftwo
 \else \expandafter \@secondoftwo
 \fi
}%
\providecommand \@ifx [1]{%
 \ifx #1\expandafter \@firstoftwo
 \else \expandafter \@secondoftwo
 \fi
}%
\providecommand \natexlab [1]{#1}%
\providecommand \enquote  [1]{``#1''}%
\providecommand \bibnamefont  [1]{#1}%
\providecommand \bibfnamefont [1]{#1}%
\providecommand \citenamefont [1]{#1}%
\providecommand \href@noop [0]{\@secondoftwo}%
\providecommand \href [0]{\begingroup \@sanitize@url \@href}%
\providecommand \@href[1]{\@@startlink{#1}\@@href}%
\providecommand \@@href[1]{\endgroup#1\@@endlink}%
\providecommand \@sanitize@url [0]{\catcode `\\12\catcode `\$12\catcode
  `\&12\catcode `\#12\catcode `\^12\catcode `\_12\catcode `\%12\relax}%
\providecommand \@@startlink[1]{}%
\providecommand \@@endlink[0]{}%
\providecommand \url  [0]{\begingroup\@sanitize@url \@url }%
\providecommand \@url [1]{\endgroup\@href {#1}{\urlprefix }}%
\providecommand \urlprefix  [0]{URL }%
\providecommand \Eprint [0]{\href }%
\providecommand \doibase [0]{https://doi.org/}%
\providecommand \selectlanguage [0]{\@gobble}%
\providecommand \bibinfo  [0]{\@secondoftwo}%
\providecommand \bibfield  [0]{\@secondoftwo}%
\providecommand \translation [1]{[#1]}%
\providecommand \BibitemOpen [0]{}%
\providecommand \bibitemStop [0]{}%
\providecommand \bibitemNoStop [0]{.\EOS\space}%
\providecommand \EOS [0]{\spacefactor3000\relax}%
\providecommand \BibitemShut  [1]{\csname bibitem#1\endcsname}%
\let\auto@bib@innerbib\@empty
\bibitem [{\citenamefont {Liu}\ and\ \citenamefont {Chen}(2024)}]{Liu24}%
  \BibitemOpen
  \bibfield  {author} {\bibinfo {author} {\bibfnamefont {H.}~\bibnamefont
  {Liu}}\ and\ \bibinfo {author} {\bibfnamefont {X.}~\bibnamefont {Chen}},\
  }\bibfield  {title} {\bibinfo {title} {Plaquette models, cellular automata,
  and measurement-induced criticality},\ }\href
  {https://doi.org/10.1088/1751-8121/ad82bc} {\bibfield  {journal} {\bibinfo
  {journal} {Journal of Physics A: Mathematical and Theoretical}\ }\textbf
  {\bibinfo {volume} {57}},\ \bibinfo {pages} {435003} (\bibinfo {year}
  {2024})}\BibitemShut {NoStop}%
\bibitem [{\citenamefont {Glauber}(1963)}]{glauber1963time}%
  \BibitemOpen
  \bibfield  {author} {\bibinfo {author} {\bibfnamefont {R.~J.}\ \bibnamefont
  {Glauber}},\ }\bibfield  {title} {\bibinfo {title} {Time-dependent statistics
  of the ising model},\ }\href@noop {} {\bibfield  {journal} {\bibinfo
  {journal} {Journal of mathematical physics}\ }\textbf {\bibinfo {volume}
  {4}},\ \bibinfo {pages} {294} (\bibinfo {year} {1963})}\BibitemShut {NoStop}%
\bibitem [{\citenamefont {Newman}\ and\ \citenamefont
  {Moore}(1999)}]{Newman_1999}%
  \BibitemOpen
  \bibfield  {author} {\bibinfo {author} {\bibfnamefont {M.~E.~J.}\
  \bibnamefont {Newman}}\ and\ \bibinfo {author} {\bibfnamefont
  {C.}~\bibnamefont {Moore}},\ }\bibfield  {title} {\bibinfo {title} {Glassy
  dynamics and aging in an exactly solvable spin model},\ }\href
  {https://doi.org/10.1103/physreve.60.5068} {\bibfield  {journal} {\bibinfo
  {journal} {Physical Review E}\ }\textbf {\bibinfo {volume} {60}},\ \bibinfo
  {pages} {5068–5072} (\bibinfo {year} {1999})}\BibitemShut {NoStop}%
\bibitem [{\citenamefont {Bortz}\ \emph {et~al.}(1975)\citenamefont {Bortz},
  \citenamefont {Kalos},\ and\ \citenamefont {Lebowitz}}]{bortz1975new}%
  \BibitemOpen
  \bibfield  {author} {\bibinfo {author} {\bibfnamefont {A.~B.}\ \bibnamefont
  {Bortz}}, \bibinfo {author} {\bibfnamefont {M.~H.}\ \bibnamefont {Kalos}},\
  and\ \bibinfo {author} {\bibfnamefont {J.~L.}\ \bibnamefont {Lebowitz}},\
  }\bibfield  {title} {\bibinfo {title} {A new algorithm for monte carlo
  simulation of ising spin systems},\ }\href@noop {} {\bibfield  {journal}
  {\bibinfo  {journal} {Journal of Computational physics}\ }\textbf {\bibinfo
  {volume} {17}},\ \bibinfo {pages} {10} (\bibinfo {year} {1975})}\BibitemShut
  {NoStop}%
\bibitem [{\citenamefont {Fenwick}(1994)}]{fenwick1994new}%
  \BibitemOpen
  \bibfield  {author} {\bibinfo {author} {\bibfnamefont {P.~M.}\ \bibnamefont
  {Fenwick}},\ }\bibfield  {title} {\bibinfo {title} {A new data structure for
  cumulative frequency tables},\ }\href@noop {} {\bibfield  {journal} {\bibinfo
   {journal} {Software: Practice and experience}\ }\textbf {\bibinfo {volume}
  {24}},\ \bibinfo {pages} {327} (\bibinfo {year} {1994})}\BibitemShut
  {NoStop}%
\bibitem [{\citenamefont {Bray}(1994)}]{bray1994theory}%
  \BibitemOpen
  \bibfield  {author} {\bibinfo {author} {\bibfnamefont {A.~J.}\ \bibnamefont
  {Bray}},\ }\bibfield  {title} {\bibinfo {title} {Theory of phase-ordering
  kinetics},\ }\href@noop {} {\bibfield  {journal} {\bibinfo  {journal}
  {Advances in Physics}\ }\textbf {\bibinfo {volume} {43}},\ \bibinfo {pages}
  {357} (\bibinfo {year} {1994})}\BibitemShut {NoStop}%
\bibitem [{\citenamefont {Aaronson}\ and\ \citenamefont
  {Gottesman}(2004)}]{Aaronson_2004}%
  \BibitemOpen
  \bibfield  {author} {\bibinfo {author} {\bibfnamefont {S.}~\bibnamefont
  {Aaronson}}\ and\ \bibinfo {author} {\bibfnamefont {D.}~\bibnamefont
  {Gottesman}},\ }\bibfield  {title} {\bibinfo {title} {Improved simulation of
  stabilizer circuits},\ }\href {https://doi.org/10.1103/PhysRevA.70.052328}
  {\bibfield  {journal} {\bibinfo  {journal} {Phys. Rev. A}\ }\textbf {\bibinfo
  {volume} {70}},\ \bibinfo {pages} {052328} (\bibinfo {year}
  {2004})}\BibitemShut {NoStop}%
\bibitem [{\citenamefont {Li}\ \emph {et~al.}(2019)\citenamefont {Li},
  \citenamefont {Chen},\ and\ \citenamefont {Fisher}}]{MIPT1}%
  \BibitemOpen
  \bibfield  {author} {\bibinfo {author} {\bibfnamefont {Y.}~\bibnamefont
  {Li}}, \bibinfo {author} {\bibfnamefont {X.}~\bibnamefont {Chen}},\ and\
  \bibinfo {author} {\bibfnamefont {M.~P.~A.}\ \bibnamefont {Fisher}},\
  }\bibfield  {title} {\bibinfo {title} {Measurement-driven entanglement
  transition in hybrid quantum circuits},\ }\href
  {https://doi.org/10.1103/PhysRevB.100.134306} {\bibfield  {journal} {\bibinfo
   {journal} {Phys. Rev. B}\ }\textbf {\bibinfo {volume} {100}},\ \bibinfo
  {pages} {134306} (\bibinfo {year} {2019})}\BibitemShut {NoStop}%
\bibitem [{\citenamefont {Coecke}\ and\ \citenamefont {Duncan}(2011)}]{ZX1}%
  \BibitemOpen
  \bibfield  {author} {\bibinfo {author} {\bibfnamefont {B.}~\bibnamefont
  {Coecke}}\ and\ \bibinfo {author} {\bibfnamefont {R.}~\bibnamefont
  {Duncan}},\ }\bibfield  {title} {\bibinfo {title} {Interacting quantum
  observables: categorical algebra and diagrammatics},\ }\href
  {https://doi.org/10.1088/1367-2630/13/4/043016} {\bibfield  {journal}
  {\bibinfo  {journal} {New Journal of Physics}\ }\textbf {\bibinfo {volume}
  {13}},\ \bibinfo {pages} {043016} (\bibinfo {year} {2011})}\BibitemShut
  {NoStop}%
\bibitem [{\citenamefont {Coecke}\ and\ \citenamefont {Kissinger}(2018)}]{ZX2}%
  \BibitemOpen
  \bibfield  {author} {\bibinfo {author} {\bibfnamefont {B.}~\bibnamefont
  {Coecke}}\ and\ \bibinfo {author} {\bibfnamefont {A.}~\bibnamefont
  {Kissinger}},\ }\bibfield  {title} {\bibinfo {title} {Picturing quantum
  processes},\ }in\ \href@noop {} {\emph {\bibinfo {booktitle} {Diagrammatic
  Representation and Inference}}},\ \bibinfo {editor} {edited by\ \bibinfo
  {editor} {\bibfnamefont {P.}~\bibnamefont {Chapman}}, \bibinfo {editor}
  {\bibfnamefont {G.}~\bibnamefont {Stapleton}}, \bibinfo {editor}
  {\bibfnamefont {A.}~\bibnamefont {Moktefi}}, \bibinfo {editor} {\bibfnamefont
  {S.}~\bibnamefont {Perez-Kriz}},\ and\ \bibinfo {editor} {\bibfnamefont
  {F.}~\bibnamefont {Bellucci}}}\ (\bibinfo  {publisher} {Springer
  International Publishing},\ \bibinfo {address} {Cham},\ \bibinfo {year}
  {2018})\ pp.\ \bibinfo {pages} {28--31}\BibitemShut {NoStop}%
\bibitem [{\citenamefont {van~de Wetering}(2020)}]{ZX3}%
  \BibitemOpen
  \bibfield  {author} {\bibinfo {author} {\bibfnamefont {J.}~\bibnamefont
  {van~de Wetering}},\ }\bibfield  {title} {\bibinfo {title} {Zx-calculus for
  the working quantum computer scientist},\ }\href@noop {} {\bibfield
  {journal} {\bibinfo  {journal} {arXiv preprint arXiv:2012.13966}\ } (\bibinfo
  {year} {2020})}\BibitemShut {NoStop}%
\bibitem [{\citenamefont {Ravindranath}\ \emph {et~al.}(2023)\citenamefont
  {Ravindranath}, \citenamefont {Han}, \citenamefont {Yang},\ and\
  \citenamefont {Chen}}]{VR_2023}%
  \BibitemOpen
  \bibfield  {author} {\bibinfo {author} {\bibfnamefont {V.}~\bibnamefont
  {Ravindranath}}, \bibinfo {author} {\bibfnamefont {Y.}~\bibnamefont {Han}},
  \bibinfo {author} {\bibfnamefont {Z.-C.}\ \bibnamefont {Yang}},\ and\
  \bibinfo {author} {\bibfnamefont {X.}~\bibnamefont {Chen}},\ }\bibfield
  {title} {\bibinfo {title} {Entanglement steering in adaptive circuits with
  feedback},\ }\href {https://doi.org/10.1103/PhysRevB.108.L041103} {\bibfield
  {journal} {\bibinfo  {journal} {Phys. Rev. B}\ }\textbf {\bibinfo {volume}
  {108}},\ \bibinfo {pages} {L041103} (\bibinfo {year} {2023})}\BibitemShut
  {NoStop}%
\bibitem [{\citenamefont {Ravindranath}\ \emph {et~al.}(2025)\citenamefont
  {Ravindranath}, \citenamefont {Yang},\ and\ \citenamefont {Chen}}]{VR_2025}%
  \BibitemOpen
  \bibfield  {author} {\bibinfo {author} {\bibfnamefont {V.}~\bibnamefont
  {Ravindranath}}, \bibinfo {author} {\bibfnamefont {Z.-C.}\ \bibnamefont
  {Yang}},\ and\ \bibinfo {author} {\bibfnamefont {X.}~\bibnamefont {Chen}},\
  }\bibfield  {title} {\bibinfo {title} {Free fermions under adaptive quantum
  dynamics},\ }\href@noop {} {\bibfield  {journal} {\bibinfo  {journal}
  {Quantum}\ }\textbf {\bibinfo {volume} {9}},\ \bibinfo {pages} {1685}
  (\bibinfo {year} {2025})}\BibitemShut {NoStop}%
\end{thebibliography}%
	\end{document}


\title{From Quantum Circuits with Ultraslow Dynamics to Classical Plaquette Models: Supplementary Material}

\maketitle
\tableofcontents
\clearpage

The central focus of this paper is the interpretation of the measurement-induced entanglement phase transition in terms of a classical transition in a system of two dimensional spins. This transition manifests as a change in the geometry of the operators that map ground states to each other.

In this supplement, we detail the methods and tools used to obtain and analyze these symmetry operators. We begin by phrasing the $N-$spin system $H_c$ as a system of equations over the binary field $\qty{0,1}^N$, and present the relevant features of the classical transition at $\beta=\infty$. We then present numerical evidence of glassy behavior in $H_c$ at finite temperature, using Monte Carlo methods.

Equipped with an understanding of a single copy of the model, we adapt these methods to study multiple copies of the model with certain boundary conditions. Specifically, we explain how the entanglement entropy of a bipartition of the quantum circuit is given by the number of independent symmetry operators that span both partitions, using elementary group theory. To extend our analysis to the finite-temperature physics of $H_c$, we utilize a generalization of the Kramers-Wannier transformation to obtain a dual model with $\sqrt{N}$ degrees of freedom.

We then turn to the fundamental question of how the classical and quantum models emerge from one another, using two distinct approaches. The first involves a cellular automaton mapping, whose utility is particularly pronounced at $\beta=\infty$. The dynamical view adopted here provides straightforward explanations of both the slow dynamics at small $p$, and the transition in the locality of symmetry operators as $p$ is tuned through $p_c$. The second avenue uses a tensor network representation of the transfer matrix to show the connection. The foremost advantage of this mapping is that it remains valid even at finite $\beta$, clearly showing how $\beta$ is related to the measurement strength, and straightforwardly provides an intuitive, graphical representation of the quantum circuit.

We conclude with additional data that underscores the ubiquity of the slow dynamics in more general circuits than the ones presented in the main text.

\section{Linear Algebra Mapping to find $G$}\label{sec:intro}

In this section, we describe an efficient process by which we obtain $G$, the group of symmetry operators that commute with $H_c$ -- this is the group of operators that transform between different ground states. Each non-trivial ground state is labelled by the locations of its nonzero bits, and so defines a multi-bit flip operator that commutes with $H_c$, generating $G$. We also discuss how this method can be modified to obtain subgroups of $G$, specifically, the boundary subgroups $G_\del$.

Classical spin models are usually phrased in terms of spin variables $m_i$ which can take values $m_i=\pm1$. General Ising interactions involving spins belonging to some set $C$ can be written as \(P^m_C\equiv\prod\limits_{m\in C} m\). Note that $P^m_C = 1$ only if an even number of the spins are in the $-1$ state, and is 1 otherwise. This fact allows us to frame these in terms of Boolean variables $s_i \equiv \frac{1+m_i}{2} \in \qty{0,1}$. Analogously, $P^m_C\to P_C =\frac{1+P^m_C}{2}$. In terms of these variables, $P_C=0$ only when an even number of $s_i$ are in the +1 state, so 
\[\prod\limits_{i\in C} m_i \to P_C\equiv \bsum s_i \]
where $\oplus$ indicates addition modulo 2.

We now frame the problem of finding the ground states of $N$ spins as a linear algebra problem over the binary field $\mF^{N}=\qty{0,1}^{N}$, as in error-correcting codes. Efficient numerical algorithms exist for these problems, allowing us to study large systems. Each term in $H_c$ -- called a ``parity check" -- is collected in the rows of a ``parity-check matrix" $H_{M\times N}$, and there are $M$ parity checks in total. For each parity check, there are 1s in the columns corresponding to the spins involved in that parity check, and 0s otherwise.

This mapping is most easily demonstrated using the classical Ising model in 1 dimension with periodic boundary conditions. Its Hamiltonian $H^I_c$ is
\[H^I_c = -\sum\limits_i m_i m_{i+1}.\]
Using the prescription to translate between $s_j$ and $m_j$, each term $m_i m_j \to 1 - 2\qty(s_i \oplus s_j)$, giving (up to overall constants)
\[
\begin{aligned}
    H^I_c &= \sum\limits_i s_i\oplus s_{i+1}
    &= |H^I\vec{s}|,
\end{aligned}\]

where $H^I$ is the parity-check matrix $|v|$ of a vector $v \in \mF^N $ denotes its Hamming norm, which is given by the number of nonzero entries in $v$. With 3 spins, for instance, the parity-check matrix is
\[H^I = \mqty(
1&1&0\\
0&1&1\\
1&0&1)\]

Matrix multiplication follows the same rules in $\mF$, so \(H_c = \sum\limits_i \qty(\bsum[j]H^I_{ij} s_j).\)

For any Hamiltonian $H_c$, a configuration $\vec{s}$ is a ground state of $H_c$ if and only if it satisfies

\begin{equation}
     H\vec{s} = 0,
     \label{eq:kern}
\end{equation}

where $s$ is an $N\times1$ binary vector. The ground states are those configurations that constitute the kernel of $H$. Let $k$ be the dimension of $\ker H$, and $\sX_j, j=1\dots k$ be its basis. Since every other state can be reached by flipping an appropriate subset of spins, we refer to any linear combination of $\sX$ as a symmetry operator (since it commutes with $H_c$) that maps $s_{x,t}=0$ to another ground state. A model with nullity $k$ thus has $2^k$ ground states, and its partition function is $\mZ = 2^k$ as well. The bit-flip operators induced by the $\sX_j$ are the generators of $G$.

Such a method is made possible because the symmetry group of $H_c$ is isomorphic to a subspace of an $N$ dimensional vector space over ${0,1}$. Each symmetry operator $\sX$ has to commute with $H_c$, resulting in $\sX$ belonging to the kernel of $H$. This structure allows us to obtain generators of the bulk and boundary symmetry groups, simply by finding an appropriate basis of $\ker H$.

Specifically, the bulk symmetry group $G_B$ is generated by those $\sX$ which leave the boundary bits unchanged. These correspond to ground states with $s_{x,t}=0$ for $t=T-1,T$, which are solutions to $H'\vec{s}=0$, where $H'$ is the matrix obtained by removing the columns corresponding to the boundary bits. $G_B$ so obtained is a subgroup of $G$. On the other hand, the boundary symmetry group is a \textit{quotient} group $G_\del \equiv G/G_B$, where operators $\sX\sim\sX'$ if their configurations on the boundary are identical $\sX|_\del = \sX'|_\del$. Equivalently, $\sX\sim\sX'$ if there exists a bulk operator $\sX_B$ such that $\sX = \sX' \sX_B$. A third interpretation of $G_\del$, from the perspective of \cref{eq:kern}, is that $G_\del$ is the set of boundary conditions $s_{x,t}; t=T-1,T$ for which a symmetry operator obeying those boundary conditions can be found.

Such a construction generalizes to any restriction of the total lattice. Thus, for any subset of bits $S$, we can define the subgroup $G_{\phi(S)}$ which consists of all $\sX$ that act trivially on the subset $S$. Formally,

\[G_{\phi(S)} = \qty{\sX\in G\text{ }|\text{ } \sX |_S = 0}.\]

This subgroup also defines a quotient group, which we term $G_{S} \equiv G/G_{\phi(S)}$, where analogously, the equivalence between two operators is given by $\sX\sim\sX'$ iff $\sX = \sX'\sX_\phi$ for some $\sX_\phi \in G_{\phi(S)}$.

\section{Symmetry Operators of $H_c$}

The symmetry operators that generate ground states from the all 0 state undergo a transition at around $p_c\approx0.25$, one manifestation of which is a change in the number of ground states changes between the two phases. However, the phenomenology of the transition is far richer -- most symmetry operators are highly non-local, 2D objects when $p<p_c$, meaning that ground states are related non-trivially.

This transition was elucidated by placing $H_c$ on a variety of different geometries in \cite{Liu24}. The order parameter for the transition is similar to the entanglement entropy of stabilizer states, in the following sense --

We showed above that the codewords/symmetry operators form a vector space $\ker H$; naturally, a basis can be chosen for $\ker H$. These basis vectors can be collected as the rows of a matrix that we call $S$. By treating this $S$ as a stabilizer that emerged from some quantum state, we can calculate the ``entanglement entropy" (and related quantities, such as ``mutual information") from $S$. The entanglement entropy (of a subset of the boundary spins) thus be calculated, goes from a ``volume-law" to an ``area-law" as $p$ is tuned through $p_c$.

Among the geometries studied in \cite{Liu24}, we focus on the case where $H_c$ is placed on a cylinder, with periodic boundary conditions in the $x$-direction. By allowing the spins at the top- and bottom-most boundaries to take any value, the automaton method generates the boundary symmetry operators $G_\del$. This allows us to directly see the change in the symmetry operators, from non-local to local, as $p$ is tuned. It is this perspective of non-locality that is crucial to explaining the entanglement phases observed in the corresponding quantum model.

\section{Markov Chain Monte Carlo (MCMC) Dynamics}
In this section, we present numerical simulations of the classical random plaquette model and provide evidence that it exhibits glassy dynamics when $p<p_c$. 

\subsection{Evolution} 

We use rejection-free continuous-time single-spin flips with Glauber (heat-bath) rates\cite{glauber1963time, Newman_1999, bortz1975new}. Let $\Delta E_i(s)$ be the energy change for flipping site $i$ in configuration $s$.
Define the per-site flip \emph{rate}
\begin{equation}
w_i(s) = \frac{1}{1+e^{\beta\,\Delta E_i(s)}} 
\end{equation}
and the total escape rate
\begin{equation}
r(s) \;=\; \sum_{i=1}^N w_i(s).
\end{equation}
The rejection-free flip rate for site $i$ is now
\begin{equation}
P_i = \frac{w_i(s)}{r(s)}.
\end{equation}

In practice, we draw $U\sim\mathrm{Unif}(0,1)$ and define $C_i=\sum_{k=1}^i P_k$ with $C_0 = 0$ and $C_N =1$ for a total of $N$ sites. Flip $s_i$ with $i$ that satisfy
\begin{equation}
C_{i - 1} < U \le C_i
\end{equation}
{ Encoding $\{C_{i = 0, \ldots, N}\}$ to a Fenwick Tree~\cite{fenwick1994new}, one can have $O(\log N)$ complexity for each Monte Carlo step if the Hamiltonian is local.}

\subsection{Clock}

While the system resides in state $s$, the probability of a Markov jump occurring in an infinitesimal interval $[t, t+\mathrm{d}t]$ is constant,
\[
\lambda \,\mathrm{d}t = r(s)\,\mathrm{d}t.
\]
Define the process
\begin{equation}
    Y_t = \exp\!\left(\int_0^t \lambda \,\mathrm{d}t\right)\,\mathbb{I}_{t < \Delta T},
\end{equation}
where $\Delta T$ denotes the waiting time to the next jump and $\mathbb{I}_{t < \Delta T}$ is the indicator of survival up to time $t$. Conditioned on past information $\mathcal{F}_t$, we have
\begin{equation}
\begin{aligned}
    \mathbb{E}[Y_{t+\mathrm{d}t}\mid \mathcal{F}_t]
    &= (1 - \lambda \mathrm{d}t)\,
       \exp\!\Bigl(\int_0^t \lambda \,\mathrm{d}t\Bigr)\,\mathbb{I}_{t < \Delta T}\,
       \exp\!\Bigl(\int_t^{t+\mathrm{d}t} \lambda \,\mathrm{d}t\Bigr)\\
    &= (1 - \lambda \mathrm{d}t)\,
       \exp\!\Bigl(\int_0^t \lambda \,\mathrm{d}t\Bigr)\,\mathbb{I}_{t < \Delta T}\,
       (1 + \lambda \mathrm{d}t)\\
    &= Y_t + \mathcal{O}(\mathrm{d}t^2).
\end{aligned}
\end{equation}
Thus $Y_t$ is a martingale, which implies
\begin{equation}\label{eq: martingale}
    \mathbb{E}[Y_t] = Y_0 = 1, \qquad \forall\,t.
\end{equation}
Writing this expectation explicitly gives
\begin{equation}
    \mathbb{E}[Y_t] = P(t < \Delta T)\, \exp\!\left(\int_0^t \lambda \,\mathrm{d}t\right),
\end{equation}
so that
\begin{equation}
    P(t < \Delta T)\, e^{\lambda t} = 1.
\end{equation}
Hence the survival probability is
\begin{equation}
    \Pr(\Delta T > t) = e^{-\lambda t},
\end{equation}
and the probability that waiting time falls in $\Delta T \in [t, t + \mathrm{d}t]$ is
\begin{equation}
    f(\Delta T = t)\mathrm{d} t = \lambda e^{-\lambda t}\mathrm{d} t.
\end{equation}
In practice, sampling is achieved by
\begin{equation}
    \Delta T = -\frac{\ln U'}{\lambda}, \qquad U'\sim \mathrm{Unif}(0,1).
\end{equation}
This construction defines the clock of the dynamics.

\subsection{Glassness}

We use Markov Chain Monte Carlo (MCMC) dynamics to investigate glassy slow-downs, as manifested in the time evolution of the internal energy. The key mechanism can be understood as follows: the characteristic relaxation time is governed by the Arrhenius law,

\begin{equation}
    T \sim \exp(\beta \Delta E),
\end{equation}
where $\Delta E$ is the characteristic energy barrier. For models where $\Delta E \sim \log L$~\cite{Newman_1999}, one immediately finds
\begin{equation}
    T \sim L^{\beta} \Rightarrow L \sim T^{1/\beta}.
\end{equation}

On the other hand, the dynamical scaling assumption posits that the excess energy density $\varepsilon(t) \equiv E(t)/N - E_0/N$ depends only on the growing correlation length $L(t)$~\cite{bray1994theory}, e.g.
\begin{equation}
    \varepsilon(t) \;\sim\; L^{-\alpha}.
\end{equation}
Combining these relations, we obtain
\begin{equation}
    \varepsilon(T) \sim g(L^\alpha) \sim g(T^{1/\beta}),
\end{equation}
with $g(\cdot)$ a universal scaling function.

\begin{figure}[ht]
  \centering
   \subfloat[\label{fig:markov_p1_non_collapse}]{%
    \includegraphics[width=0.35\textwidth]{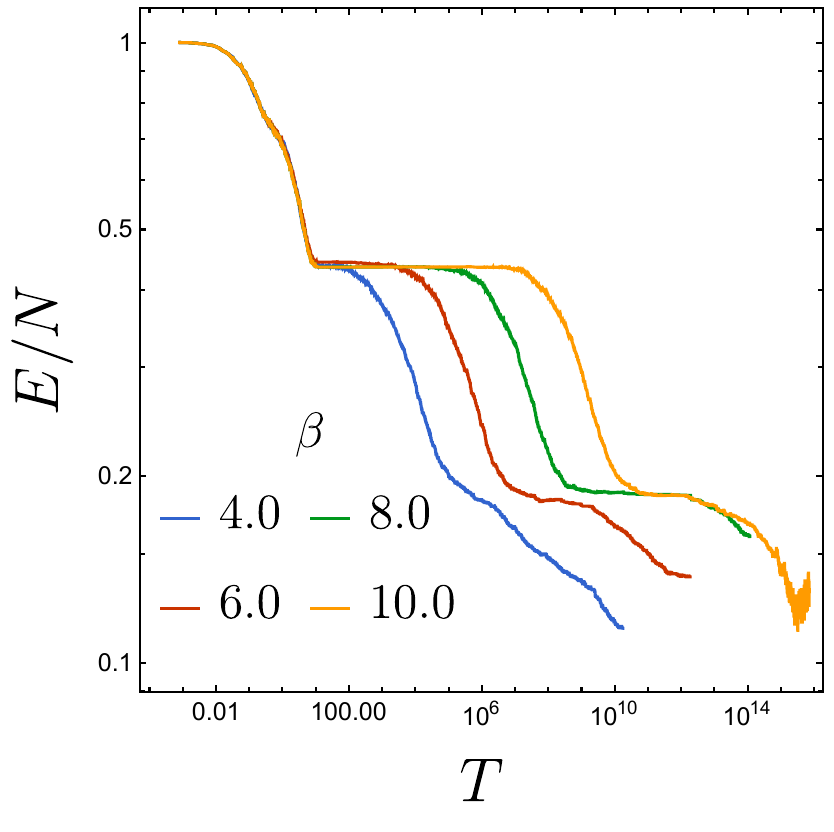}%
  }\quad
  \subfloat[\label{fig:markov_p1_collapse}]{%
    \includegraphics[width=0.35\textwidth]{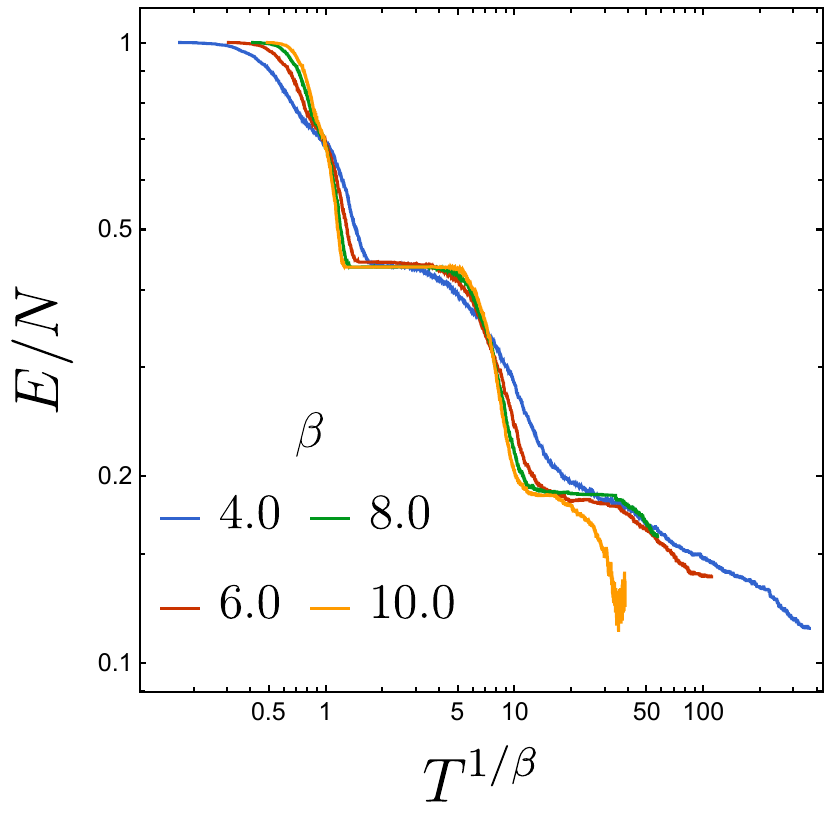}%
  }
  \caption{\label{fig: p0_glassy_slow_down} Data collapse of Continuous Time MCMC evolution of the internal energy of random plaquette model with $p = 0$, $N = 128 \times 128$.}
\end{figure}

We consider an \(L\times L\) random plaquette model on a torus with Hamiltonian, which is the same classical Hamiltonian as shown in the main text,
\begin{equation}
    H[s] \;= \;\sum_i \bigl(1-h_i\bigr)
\end{equation}
where the local term \(h_i\) is random: with probability \(1-p\),
\begin{equation}
    h_i \;=\; s_i \prod_{j\in \text{Nbr}_i} s_j,
\end{equation}
with \(\text{Nbr}_i\) the neighbors of site \(i\), and otherwise
\begin{equation}
    h_i \;=\; s_i.
\end{equation}
We initialize with i.i.d.\ spins \(s_i=\pm1\) with equal probability and track the energy density \(E(T)= 1/N H[s(T)]\, , N = L^2\) under the continuous-time dynamics described earlier. As shown in Fig.~\ref{fig: p0_glassy_slow_down} (a), in the clean limit with $p=0$, the time evolution exhibits a sequence of plateaus.

This behavior is analogous to that discussed in Ref.~\cite{Newman_1999} and can be understood as follows. Initially, the system contains an extensive number of defects, defined as $h_i$ that violate the ground state constraint. Under MCMC dynamics, a single defect cannot vanish on its own. It can move around the lattice, but annihilation is only possible when five defects come together as neighbors. The relaxation proceeds hierarchically: nearby defects merge and annihilate first; defects at larger separations then gradually come together and annihilate; and finally, even defects separated by distances of order the system size are eliminated. Only after this last stage does the system fully relax to the ground state. This hierarchical dynamics gives rise to the sequence of plateaus observed in the simulations and captures the glassy nature of the model.

The lifetime of the plateau can be estimated as follows: If the minimal barrier to eliminate defects of linear distance $D$ scales as \( \Delta E(D)\sim \log D\), then by Arrhenius activation the typical lifetime obeys
\begin{equation}
    \tau_D \;\sim\; \exp\!\big(\beta\,\Delta E(D)\big) \;\sim\; D^{\beta}.
\end{equation}
 Consequently, for time windows $T \ll \tau_D$ a distance-$D$ defect is effectively frozen, and the observed energy decay is dominated by annihilation of much smaller defects ($d\ll D$). This produces an extended energy plateau. When $T \sim \tau_D$, a rare cooperative event removes the distance-$D$ defect and the energy drops to a new level. The annihilation of the defects continues until $d\sim L$ defects annihilate and $E\to 0$. As shown in Fig.~\ref{fig: p0_glassy_slow_down} (b), this hierarchical dynamics is further supported by the data collapse obtained when curves for different \(\beta\) are plotted against \(T^{1/\beta}\), consistent with a logarithmic barrier,  
\[
    \Delta E \sim \log L .
\]  

\begin{figure}
  \centering
   \subfloat[\label{fig:markov_p1_non_collapse} $\beta = 4$]{%
    \includegraphics[width=0.35\textwidth]{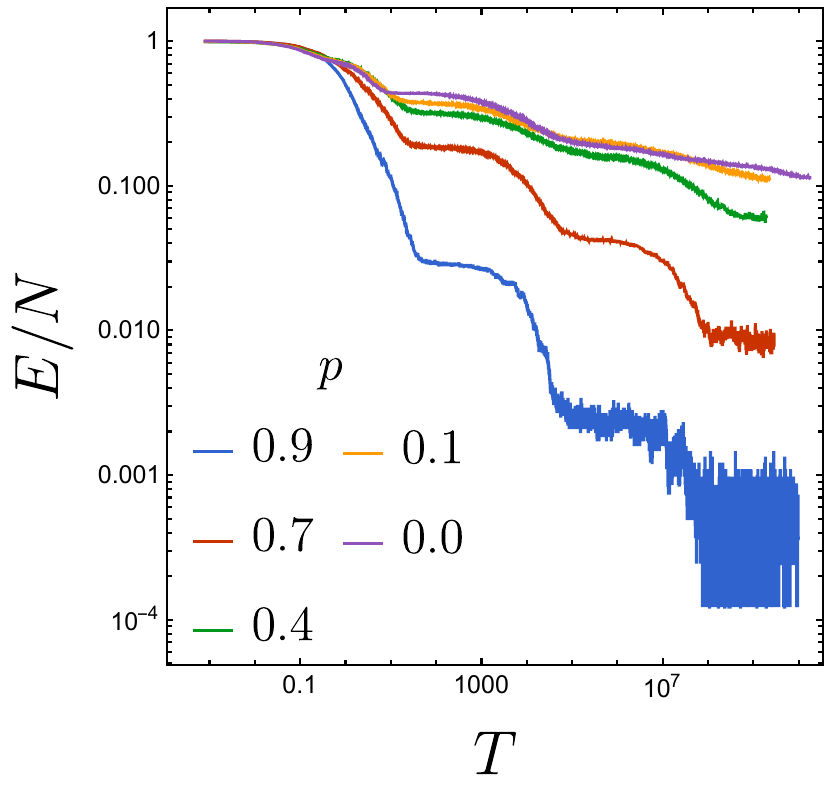}%
  }\quad
  \subfloat[\label{fig:markov_p1_collapse}$\beta = 6$]{%
    \includegraphics[width=0.35\textwidth]{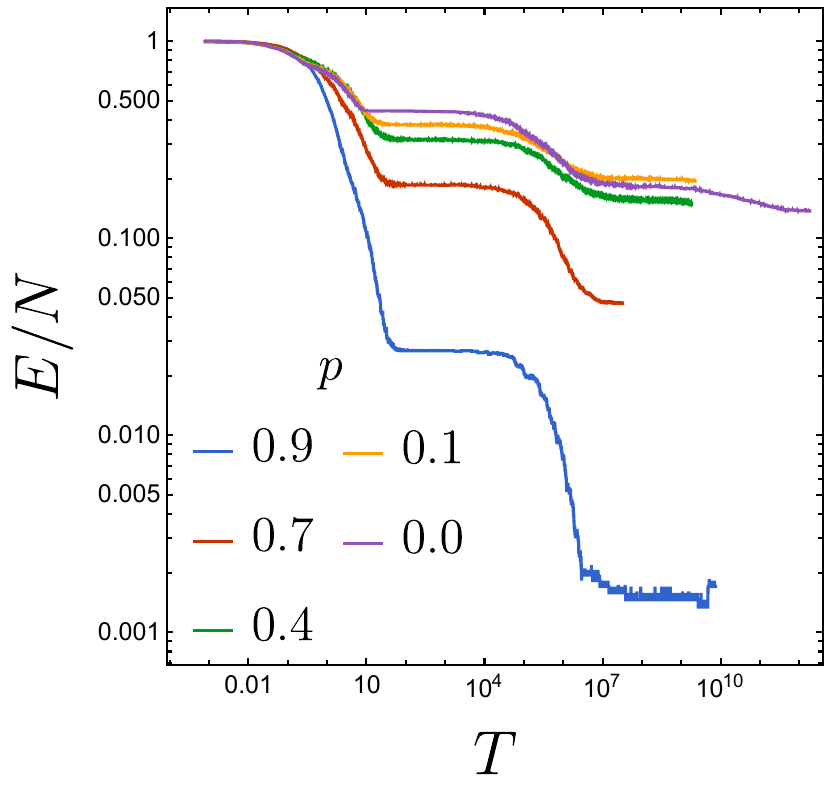}%
  }
  \caption{\label{fig: p_glassy_slow_down} Continuous Time MCMC evolution of the internal energy of random plaquette model with $p = 0.0, 0.1, 0.4, 0.7, 0.9$, $\beta = 4, 6$, $N = 128 \times 128$.}
\end{figure}

\begin{figure}[ht]
  \centering
  \subfloat[\label{fig:markov_p1_un} $p = 0$]{%
    \includegraphics[width=0.35\textwidth]{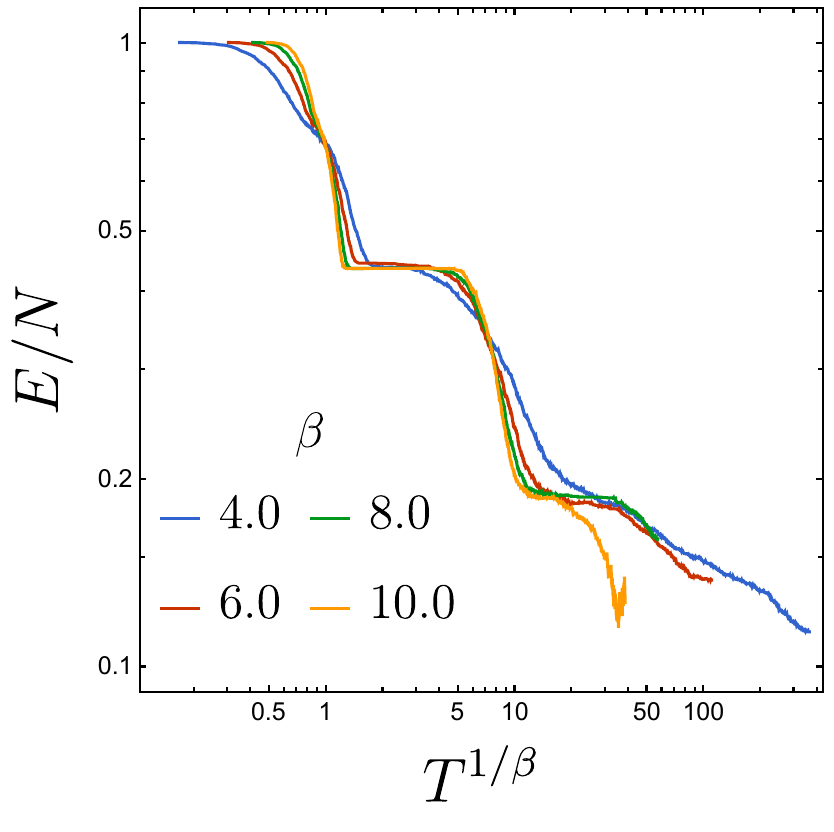}%
  }\quad
  \subfloat[\label{fig:markov_p0.9_un} $p = 0.1$]{%
    \includegraphics[width=0.35\textwidth]{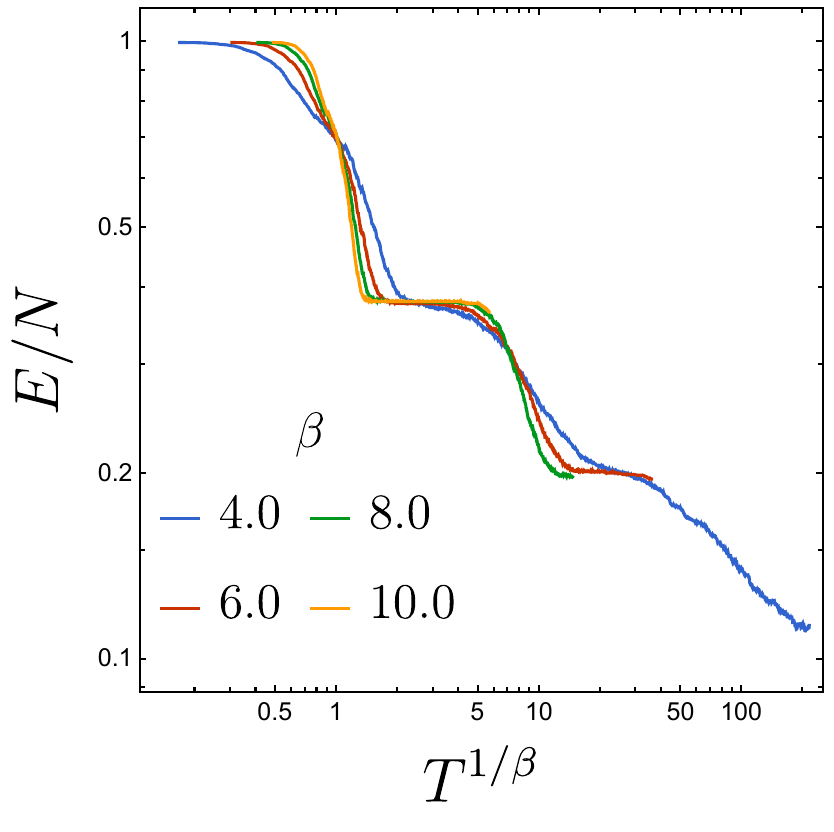}%
  }\quad
   \subfloat[\label{fig:markov_p0.6_un} $p = 0.4$]{%
    \includegraphics[width=0.35\textwidth]{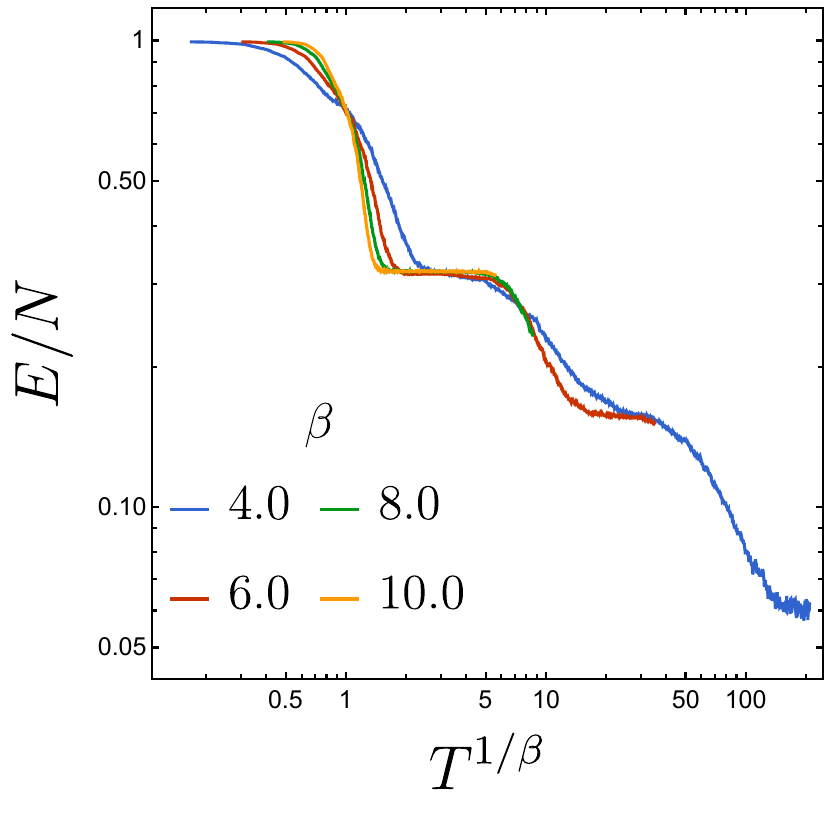}%
  }\quad
  \subfloat[\label{fig:markov_p0.3_un} $p = 0.7$]{%
    \includegraphics[width=0.35\textwidth]{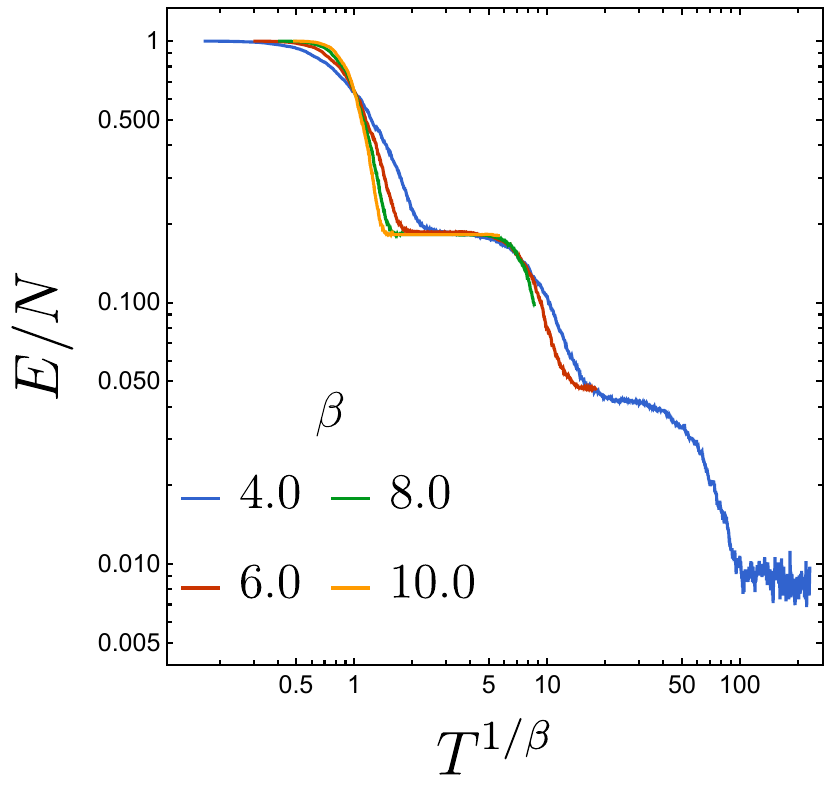}%
  }
  \caption{\label{fig: internal_energy_collapse} Continuous time MCMC evolution of the internal energy of random plaquette model with $N = 128 \times 128$.}
\end{figure}

As we move away from the \(p=0\) limit, extended energy plateaus persist, as shown in Fig.~\ref{fig: p_glassy_slow_down}, with the energy failing to relax to zero even over very long times. Although the precise form of the energy barrier in this regime is not fully understood, we adopt the same scaling form used for \(p=0\), and remarkably, the curves for different parameters still collapse onto a single line as shown in Fig.~\ref{fig: internal_energy_collapse}. The persistent nonzero energy further reflects the slow, glassy dynamics of the model. 

For \(p>p_c\), numerical simulations show initial plateaus associated with the relaxation of defects at \(O(1)\) distances. Beyond this early stage, however, the system rapidly relaxes to \(E \to 0\), and no additional plateaus are observed.

\section{Symmetry groups and Entanglement}
\label{sec:sym-ent}

In the main text, we had shown that the entanglement entropy $\se$ was given by the difference between $2k^{(2)}$ and $k^{(4)}$, using the replica trick. This relies on an alternate definition of $\se$, defined on two copies of the density matrix $\rho$
\[\se = -\log \tr\qty(\rho\otimes\rho SW_A),\]
where $SW_A$ is an operator that swaps the $A$ qubits \textit{between} copies. The equivalence between this definition and $\se\equiv -\log \tr \rho^2_A$ can most easily be shown graphically. Any density matrix $\rho$ can be written in the form
\[\rho =\sum\limits_{n,n'} \rho^{n'_A,n'_{\sA}}_{n_A,n_{\sA}}\ket{{n'_A,n'_{\sA}}}\bra{{n_A,n_{\sA}}},\]
with each $n_{A/\sA}$ labelling a complete basis on subsystem $A (\sA)$. The reduced density matrix $\rho_A$ is obtained by tracing over the $\sA$ states. Graphically, this process is represented by

\begin{equation}
	\includegraphics{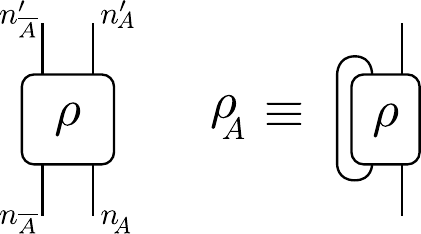}
\end{equation}
In these diagrams, as in all tensor network diagrams, two joined lines denote a ``contraction", or a summation over the shared, connected indices. The result of any such diagram depends only on its topology, so different parts of the diagram can be deformed arbitrarily, provided no lines are cut. With these two facts, we can manipulate the diagram that defines $\tr \rho^2_A$, to obtain

\begin{equation}
	\includegraphics{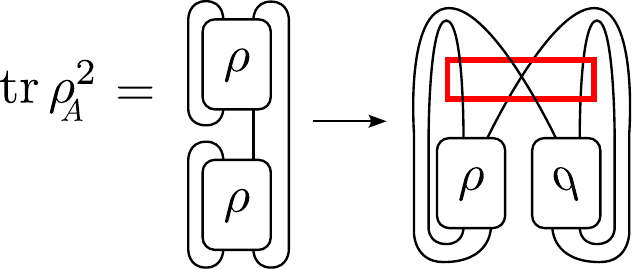}
\end{equation}

The diagram to the right of the arrow is obtained by reflecting the $\rho$ tensor on the top and bringing it beside the bottom $\rho$ tensor. Note that none of the connectivities have been altered in this process. The part of the diagram enclosed in red swaps one set of the $A$ indices between the copies, while leaving the $\sA$ indices intact. However, we can equivalently treat the red box as an operator $O$ that acts on the doubled Hilbert space. Therefore, $\tr \rho^2_A$ reduces to the expectation value of $O$ on $\rho\otimes\rho$. Finally, we have
\begin{equation}
	\includegraphics{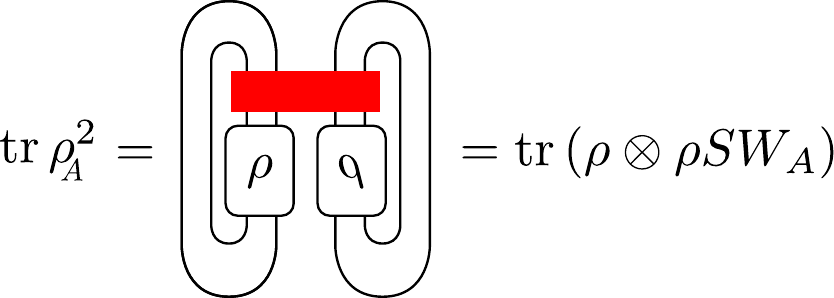}
\end{equation}
where, as above, $SW_A$ swaps $A$ qubits between the copies. In our work, we do not normalize $\rho$ after each measurement. Therefore, the complete expression for $\se$ is

\begin{equation}
	2^{-{\se}} = \frac{\tr \rho^2_A}{\qty(\tr\rho)^2} = \frac{\tr\qty(\rho\otimes\rho SW_A)}{\tr\rho \tr\rho} = \frac{\bra{\psi}\bra{\psi}SW_A\ket{\psi}\ket{\psi}}{\braket{\psi}\braket{\psi}}.
\end{equation}

It is this $SW_A$ operator that leads to the boundary conditions required to calculate $\mZf$. For clarity, we rewrite the numerator as $\bra{\psi_1}\bra{\psi_2}SW_A\ket{\psi_3}\ket{\psi_4}$. Since the unnormalized wavefunctions can be written as a superposition $\ket{\psi} = \sum\limits_s \mZ(s)\ket{s}$, $SW_A$ requires the $A$ subsystem of 1 (2) to be connected to that of 4 (3), while the $\sA$ subsystems of 1 (2) are connected to 3 (4). Identifying each copy of $\psi$ as an independent copy of $\mZ$, we arrive at the expression in the main text
\begin{equation}
	\se = -\log \frac{\mZf}{{\mZt}^2} .
\end{equation}
The normalization factor $\mZt\equiv\braket{\psi}$ follows analogously, requiring the boundaries to be connected uniformly.

Recall, $k^{(n=2,4)}$ denotes the number of independent codewords of $n$ copies with their respective boundary conditions.  We now concretely show that these $k$ are directly related to the sizes of different boundary symmetry groups of a \textit{single} copy of the model. In what follows, the boundary is labelled by $\del = \qty{(x,t) | t= T-1, T}$.
	
We begin with $k^{(2)}$, by finding symmetry operators of two copies of the system constrained to have the same boundary. Note that any \sX of $n$ copies of the system has to be constructed from the $\sX$ of single copies (otherwise, the bulk parity checks will remain unfulfilled). In equations,

\begin{equation}
    k^{(2)} = \log \sum N^1_{s_\del} N^2_{s_\del},
    \label{eq:k2}
\end{equation}
where $N^i_{s_\del}$ is the total number of symmetry operators on copy $i$, with a boundary configuration $\qty{s_\del }$. In the calculations that follow, we utilize the following definition of ``boundary subgroups". For any region $A\subseteq\del$, 
\[|G_{]A[}| = \qty{ \sX \in G | \sX|_A = 0},\]
i.e., $|G_{]A[}|$ is the subgroup of symmetry operators that exclusively act \textit{outside} $A$. For convenience, we additionally define the bulk symmetry group $G_B \equiv |G_{]\del[}| $ which act only in the bulk of a given copy. The sum can then be written as a summation over all symmetry operators as

\begin{equation}
    \begin{aligned}
        \sum N^1_{s_\del} N^2_{s_\del} &= \sum\limits_{s_\del}\sum\limits_{\sX,\sX'\in G} \delta\qty(X|_\del = s_\del)\delta\qty(X'|_\del = s_\del)\\
        &= \sum\limits_{\sX\in G} \qty(\sum\limits_{s_\del}\delta\qty(\sX|_\del = s_\del))\sum\limits_{\sY\in G_{]\del[}}\\
        &= |G||G_{]\del[}|.
    \end{aligned}
\end{equation}

The second line can be explained as follows. Consider two symmetry operators $\sX$ and $\sX'$ which are equal on $\del$. Their combined operator $\sX\sX'$ is clearly a bulk operator, since $\sX\sX' |_\del = 0$. Thus, given an $\sX$ with a specific configuration $s_\del$, all other $\sX'$ with the same boundary configuration can be reached by acting the bulk operators in $G_B$ on $\sX$, for each $\sX$. This reasoning holds for any subset of $\del$ as well, particularly for the subsystems $A$ and $\sA$.

\[k^{(2)} \equiv \log(|G||G_B|)\]

As for $k^{(4)}$, the BCs on $A$ and $\sA$ need to be specified separately. $A$ and $\sA$ denote regions at the boundary ($t=T-1, T$); $A$ contains sites with $x=1\dots\frac{L}{2}$, while $\sA$ is its complement (on the boundary). $s_S$ denotes the spin configuration on set $S$. From the replica picture, with the replica indices $i$ dropped,

\[k^{(4)} = \log \sum\limits_{\substack{s_A, s'_A\\ s_{\sA}, s'_{\sA} }} N_{s_A s_{\sA}} N_{s_A s'_{\sA}} N_{s'_A s_{\sA}} N_{s'_A s'_{\sA}},\]

We focus on a part of the summation, specifically, the terms $\sum\limits_{s_A} N_{s_{A} s_{\sA}}N_{s_{A} s'_{\sA}}$, which can be rewritten as a summation over $G$ as

\begin{equation}
    \begin{aligned}
    \sum_{s_A}\sum\limits_{\sX,\sX'} \delta\qty(X|_\del = s_A s_{\sA}) \delta\qty(X|_\del = s_A s'_{\sA})\\
    =\sum\limits_{\sX\in G} \qty(\sum\limits_{s_{A}}\delta\qty(\sX|_\del = s_As_{\sA}))\sum\limits_{\sY\in G_{]A[}} \delta\qty(\sY|_{\sA} = s_{\sA} \bsum[] s'_{\sA})\\
    =\sum\limits_{\sX} \delta\qty(\sX|_{\sA} = s_{\sA}) \sum\limits_{\sY\in G_{]A[}} \delta\qty(\sY|_{\sA} = s_{\sA} \bsum[] s'_{\sA})
    \end{aligned}
\end{equation}

At this point, to simplify notation, we introduce the following shorthands

\begin{equation}
    \begin{aligned}
        \delta_{s_{\sA}} &\equiv \delta\qty(\sX|_{\sA} = s_{\sA}),\\
        \delta'_{s_{\sA}} &\equiv \delta\qty(\sX'|_{\sA} = s_{\sA}),\\
        \delta^Y_{s_{\sA}} &\equiv \delta\qty(\sY|_{\sA} = s_{\sA}).
    \end{aligned}
\end{equation}

Now, incorporating the terms from $\sum\limits_{s'_A} N_{s_{A} s_{\sA}}N_{s_{A} s'_{\sA}}$ (notice the priming (') on $s'_A$),

\begin{equation}
    \begin{aligned}
        k^{(4)} = \log \sum_{s_{\sA}, s'_{\sA}} \sum\limits_{\substack{\sX, \sX'\in G\\ \sY,\sY'\in G_{]A[}}} \delta_{s_{\sA}} \delta'_{s_{\sA}} \delta^Y_{s_{\sA}\bsum[]s'_{\sA}} \delta^{Y'}_{s_{\sA}\bsum[]s'_{\sA}}\\
        = \log \sum_{s_{\sA}, s'_{\sA}} \sum\limits_{\substack{\sX, \sX'\in G\\ \sY,\sY'\in G_{]A[}}} \delta_{s_{\sA}} \delta'_{s_{\sA}} \delta^Y_{s'_{\sA}} \delta^{Y'}_{s'_{\sA}},
    \end{aligned}
\end{equation}

where the last line follows from redefining the sum over $s'_{\sA}$ to a sum over $s_{\sA}\bsum[]s'_{\sA}$. Since both $s$ and $s'$ are dummy variables, this does not change $k^{(4)}$. The sums involving $\sX$ and $\sY$ are now disconnected, and can be performed separately. Summing over $\sX$ and $\sX'$, we have

\begin{equation}
    \begin{aligned}
        \sum\limits_{s_{\sA}}\sum\limits_{\sX,\sX'} \delta_{s_{\sA}} \delta'_{s_{\sA}} =\sum\limits_{s_{\sA}}\sum\limits_{\sX} \delta_{s_{\sA}} \sum\limits_{\sY\in G_{]\sA[}}\\
        =\sum\limits_{\sX\in G} \qty(\sum\limits_{s_{\sA}} \delta_{s_{\sA}}) \sum\limits_{\sY\in G_{]\sA[}}\\
        = |G||G_{]\sA[}|,
    \end{aligned}
\end{equation}
while summing over $\sY$ and $\sY'$ leaves us with
\begin{equation}
    \begin{aligned}
        \sum\limits_{s'_{\sA}}\sum\limits_{\sY,\sY'\in G_{]A[} } \delta^Y_{s'_{\sA}} \delta^{Y'}_{s'_{\sA}} &= \sum\limits_{s'_{\sA}}\sum\limits_{\sY} \delta^Y_{s'_{\sA}} \sum\limits_{\sY'\in G_{]A[}} \delta^{Y'}_{0}\\
        &=\sum\limits_{\sY\in G_{]A[}} \qty(\sum\limits_{s_{\sA}} \delta_{s_{\sA}}) \sum\limits_{\sY'\in G_B}\\
        &= |G_B||G_{]A[}|.
    \end{aligned}
\end{equation}

The second line follows from the fact that

\begin{gather}
    \sum\limits_{\sY'\in G_{]A[}} = \sum\limits_{\sY'\in G} \delta\qty(\sY'|_{A} = 0)\nonumber\\
        \implies\sum\limits_{\sY'\in G_{]A[}} \delta^{Y'}_{0} = \sum\limits_{\sY'\in G} \delta\qty(\sY'|_{A} = 0) \delta\qty(\sY'|_{\sA} = 0) = \sum\limits_{\sY'\in G_B} = |G_B|.
\end{gather}

Putting these equations together, we have 	
\begin{equation}
    \se = \log |G||G_B| - \log |G_{]A[}| - \log |G_{]\sA[}|.
\end{equation}

Notice that for any part of the boundary $S$ and its subset $S'\subseteq S\subseteq \del$, 
\[G_{]\del[}\equiv G_B \subseteq G_{]X[} \subseteq G_{]X'[}.\]
Following this, we can rewrite
\begin{equation}
    |\mathcal{G}| = |\mathcal{G}/G_B||G_B|,
\end{equation}
for $\mathcal{G} = G_{]A[}, G_{]\sA[}, G$. The quotient groups -- consider $G_{]A[}/G_B$, for instance -- can be interpreted as a set\footnote{Moreover, it is a group of bitstrings $s_{\sA}$ under element-wise $\bsum[]$.} of boundary configurations on $\sA$ labelled $s_{\sA}$ that admit an $\sX \in G$, with the bits on $A$ set to 0. $|G_{]A[}/G_B|$ thus counts the number of symmetry operators (modulo bulk configurations) that are supported completely outside $A$. Following cancellations of factors of $|G_B|$, we have

\begin{equation}
    \se = \log |G/G_B| - \log |G_{]A[}/G_B| - \log |G_{]\sA[}/G_B|.
    \label{eq:ee_from_rep}
\end{equation}

The origin of the entanglement phase transition at zero temperature is now evident. In the volume law phase $p<p_c$, most symmetry operators are nonlocal, irreducibly spanning both $A$ and $\sA$, so that few symmetry operators are completely 0 on either $A$ or $\sA$. However, when $p>p_c$, the symmetry operators are local, point-like objects localized near the boundary; nearly every $\sX\in G$ can be chosen to act completely outside either $A$ or $\sA$. The expression in \cref{eq:ee_from_rep} intricately links the non-locality of symmetry operators to the volume law phase, thus complementing the intuitive geometric picture of the MIPT presented in this work.

\subsection{Finite Temperature Transition: A Free Energy Argument}\label{sec:free_energy}

Although the proof presented above is exact, we can gain more clarity as to \textit{why} the entanglement entropy is a difference in the number of symmetry operators by enumerating the contributions to $\mZf$ and $\mZt$ explicitly.	The calculation of $\se$ involves 2 factors of $\mZt$ and one of $\mZf$, as can be seen from its definition
\[\se = -\log \frac{\mZf}{{\mZt}^2}.\]
Therefore, both ${\mZt}^2$ and $\mZf$ are partition functions on 4 copies of $H_c$, albeit with different BCs. Labelling the copies 1-4, one factor of $\mZt$ involves identifying the boundaries of copies 1 and 2. Similarly, the other factor of $\mZt$ requires identifying those of copies 3 and 4.

As mentioned earlier, the zero temperature contributions can only be composed of $\sX$ of an individual $H_c$. Since operators from $G_B$ act independently in the bulk and do not alter the BCs, we will restrict our attention to operators from $G/G_B$ (recall from the previous section that $G/G_B$ consists of BCs $s_\del$ that can admit a symmetry operator). In this section, we will move away from the $\beta=\infty$ limit by considering the highest weight contributions to $\mZt$ and $\mZf$.

\begin{figure}
    \centering
    \includegraphics[width=0.35\linewidth]{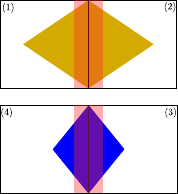}
    \hspace{0.2\linewidth}\includegraphics[width=0.4\linewidth]{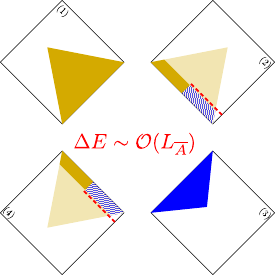}
    \caption{The boundary excitations that ultimately lead to a thermal phase transition in \se. (Left) The configurations all have the same weight in 2 copies. (Right) Placing the $\sX$ on 4 copies leads to an energy cost $\Delta E \sim L$. The shaded red region denotes the shared boundaries.}
    \label{fig:excite_arg}
\end{figure}

To set notation, $s^a_A$ denotes the boundary configuration on $A$ (likewise for $\sA$) on the copy numbered $a$, while $s^a_\del$ denotes the BC on the entire boundary. We set the ground state energy $E_0$ of $H_c$ to be 0. We choose BCs $s^{1/3}_\del$ independently, since there are no (direct) constraints between the boundaries of 1 and 3, either in the calculation of $\mZt$ or $\mZf$. For any choice of $s^1_\del$, the bulk of copies 1 and 2 can be placed in a configuration corresponding to a symmetry operator, ensuring that this contributes a weight of $e^{-\beta E_0}\equiv 1$ to $\mZt$ (\cref{fig:excite_arg}, left).

In $\mZf$, not every independently chosen configuration on 1 and 3 has the same weight. A configuration so chosen contributes a weight 1 to $\mZf$ only if BCs $s^1_A s^3_{\sA}$ \textit{and} $s^3_A s^1_{\sA}$ also admit $\sX$s. This explains why, at $\beta=\infty$, $\mZf$ only admits those configurations composed of $\sX$ that act completely outside $A$ or $\sA$, so that the BC $s^3_{\sA}$ has no bearing on $s^1_A$ (and vice-versa). These discarded configurations generally lead to an energy penalty of $\Delta E = \mO(\min(L_A, L_{\sA}))$, as shown in \cref{fig:excite_arg} (right) . For a BC $s^2_\del = s^1_A s^3_{\sA}$ that does \textit{not} admit an $\sX$, a possible minimum energy configuration can be obtained by constructing the $\sX$ corresponding to $s^1_\del$ on copy 2, and then flipping the bits $j_{\sA}$ in $\sA$ for which $s^1_{j_{\sA}}\neq s^3_{j_{\sA}}$. An energy penalty of 1 is incurred for each bit that was flipped on the boundary of a ground state. Equivalently, one can instead begin by constructing an $\sX$ on copy 2 from the BC $s^3_\del$, and flip the spins on $A$. Finally, we choose the configuration which requires the fewest number of bit flips. However, the extensive difference between BCs of different $\sX$ in $G/G_B$ means that a finite fraction of $\sA$ or $A$ needs to be flipped on copy 2, leading to the energy penalty $\Delta E = \mO(\min(L_A, L_{\sA}))$. The weight of these configurations goes as 
\[e^{-\beta\Delta F}\sim e^{-\beta\min(L_A, L_{\sA})} 2^{\gamma(p) L},\]since the number symmetry operators that cannot be isolated on either $A$ or $\sA$ grows exponentially in $L$, \textit{provided} $p<p_c$. Moreover, since $L_A \equiv \alpha L$ is chosen to be a finite fraction of $L$, the free energy difference $\Delta F$ between these disordered configurations and the ground states can be expressed as
\[\Delta F \sim \min(\alpha,1-\alpha)L - (1/\beta)\gamma(p)\log 2 L.\]

Thus, at a critical $T_c$	
\begin{equation}
    1/\beta_c\sim \frac{\min(\alpha,1-\alpha)}{\gamma(p)\log 2},
\end{equation}
we expect a thermal phase transition in $\frac{\mZf}{{\mZt}^2}$, which is exactly the weak-measurement entanglement phase transition observed in our quantum circuit. This restriction to boundary excitations alone is justified by the fact that our model consists only of local interactions, and that away from the boundary, the contributions to $\mZf$ and ${\mZt}^2$ are identical.

It is interesting to note that without fixed BCs, a thermal phase transition cannot occur since large excitations can be introduced with a low energy cost by flipping the boundary spins appropriately. However, once different copies are forced to have partially identical boundaries, as in the replica picture, these previously low-energy excitations gain an energy proportional to $L$. This in turn allows for a competition between the energetic ($\Delta E$) and entropic ($| (G/G_B) \ (G_{]A[}/G_B)$) factors, thereby hosting a finite temperature phase transition.

\section{Generalized Kramers-Wannier Transformation}

The nonlocality of the $\sX$s provides an illuminating and geometric lens through which to understand the measurement-induced phase transition. Unfortunately, it no longer applies away from $\beta=\infty$. At finite temperatures, one needs to account for excitations, which generally involves grappling with all $N$ spins. Since much of the relevant physics occurs at the boundary of the model, it would be highly desirable to obtain a description in terms of $\mO(L\sim\sqrt{N})$ variables. Fortunately, a well-studied generalization of the familiar Kramers-Wannier transformation can help us achieve this goal.

The thermodynamics of a generic parity-check matrix $H$ can be equivalently phrased in terms of the \textit{redundancies} between parity checks. A model is said to have $Q$ redundancies if there exist linearly independent vectors $R_1,R_2,\dots,R_Q$ over $\mF^M$, such that

\NewDocumentCommand{\mrule}{}{\ensuremath{\rule[0.5ex]{0.5cm}{0.4pt}}}

\[ \mqty(%
    \mrule&R_1&\mrule\\
    \mrule&R_2&\mrule\\
    &\vdots&\\
    \mrule&R_Q&\mrule\\) H = 0\]

We collectively refer to the matrix of $R_j$s as $R$. We now present a derivation of the transformation, after which its utility will become clear.

We begin by attempting to calculate the partition function of $H$, phrased in terms of spin variables. We denote each row (parity check) of $H$ as $C_i$ for $i=1,\dots,M$. The partition function is given by 
\begin{equation}
    \mZ_H(\beta) = \sum\limits_{s_j=\pm1} \prod\limits_{i=1}^M \exp(\beta\prod\limits_{j\in C_i} s_j).
\end{equation}

A spin $j$ is ``in" the parity check $C_i$ when $H_{ij}=1$. As a first step, for any $\theta$ such that $\theta^2=1$, note that
\[\exp(\beta\theta) = \cosh(\beta)\qty(1 + \alpha \theta),\]
where $\alpha\equiv \tanh(\beta)$. With some abuse of notation, we also define $C_i\equiv \prod\limits_{j: H_{ij} = 1} s_j$. We then obtain

\begin{equation}
    \begin{aligned}
        \mZ_H(\beta) &= (\cosh(\beta))^M\sum\limits_{s_j=\pm1} \prod\limits_{i=1}^M \qty(1 + \alpha C_i)\\
        &= \cosh(\beta)^M\sum\limits_{n_k=0,1}\sum\limits_{s_j=\pm1} (\alpha C_1)^{n_1}\dots(\alpha C_M)^{n_M}
    \end{aligned}
\end{equation}

By virtue of the summation of $\qty{s_j}$ over $\pm1$, the term
\[\sum\limits_{s_j=\pm1} (\alpha C_1)^{n_1}\dots(\alpha C_M)^{n_M}\]
is non-zero only when the product of parity checks involves no spins. Explicitly,
\[
\begin{aligned}
    \sum\limits_{s_j=\pm1} (\alpha C_1)^{n_1}\dots(\alpha C_M)^{n_M} \neq 0\\
    \implies (\alpha C_1)^{n_1}\dots(\alpha C_M)^{n_M} = 1\\
    \implies \vec{n}^TH = 0
\end{aligned}
\]
Additionally, when the product involves no spins, the trivial sum over $N$ spins leads to an overall prefactor of $2^N$.

\begin{equation}
    \begin{aligned}
        \mZ_H(\beta) &= 2^N\cosh(\beta)^M\sum\limits_{n_k}\alpha^{\sum\limits_{i=1}^{M} n_i} \delta(\vec{n}^TH = 0)\\
        &=2^N\cosh(\beta)^M \mZ_{R}(\alpha),
    \end{aligned}
\end{equation}
where we have defined a new partition function 
\[\mZ_{R}(\alpha) = \sum\limits_{n_k}\alpha^{\sum\limits_{i=1}^{M} n_i} \delta(\vec{n}^TH = 0).\] It is clear that we only sum over the vectors $\vec{n}$ that belong to the left kernel of $H$, $\ker_LH$. A potentially more efficient way to perform this sum is to choose a basis of $\ker_LH$, and enumerate the sum over all linear combinations of those basis vectors.

The redundancies $R_j$ provide \textit{exactly} such a basis. Thus, every $\vec{n}\in\ker_LH$ can be written in the form $R_T\vec{q}$, where $\vec{q} \in \mF^Q$ collects the $Q$ new degrees of freedom $\qty{q_j}$. Moreover, the exponent of $\alpha$, \({\sum\limits_{i=1}^{M} n_i}\), is the Hamming distance of $\vec{n}$, $|\vec{n}|$, ultimately giving

\begin{equation}
    \mZ_R(\alpha) = \sum\limits_{\vec{q}}\alpha^{|R^T \vec{q}|}.
\end{equation}

Since $\alpha\leq1$, each non-zero entry in $R^T \vec{q}$ imposes an energy penalty in the following sense -- treating $R^T$ as an $M\times Q$ parity check matrix, the weight of a configuration is reduced by the number of violated parity checks. Each row of $R^T$ can be labelled by a parity check of $H$, with $R^T_{ij} = 1$ if the redundancy $R_j$ involves $C_i$.

\def\tC{\ensuremath{\widetilde{C}}}

For completeness, we rewrite $\mZ_R$ in terms of spin-variables $p_j = 1 - 2q_j$, analogous to the $m$ and $s$ variables of Sec.~\ref{sec:intro}. Each row of $R^T$ constitutes a parity check, which we label $\tC_i$. Recall that
\[\bsum[j] R^T_{ij}q_j = 1 - \prod\limits_{j\in\tC_i} p_j, \]
where, as before, a spin $j$ belongs to parity check $\tC_i$ if $R^T_{ij} = 1$, and not otherwise. In terms of $p_j$, we have

\begin{equation}
\begin{aligned}
    \mZ_R(\alpha) = \sum\limits_{\qty{p_j=\pm1}}\alpha^{1-\sum\limits_i\prod\limits_{j\in\tC_i} p_j}\\
    = \sum\limits_{\qty{p_j=\pm1}} \exp(-\beta'H_R),
\end{aligned}
\end{equation}

where the dual Hamiltonian $H_R$ is 
\[H_R\equiv-\sum\limits_i\prod\limits_{j\in\tC_i} p_j.\]

Finally, by comparing the coefficients for a given parity check, we relate the partition function of $H_c$ at temperature $\beta$, to that of $H_R$ at $\beta'$ with

\[\exp(-2\beta') = \tanh(\beta),\]

which is familiar from the Kramers-Wannier duality.\footnote{In the case of the Ising model, $H_R$ is in fact equal to $H_c$, which directly demonstrates the Kramers-Wannier duality at $\beta_c$, avoiding perturbative high- and low-temperature expansions.}

Additionally, this mapping also explains the absence of a phase transition when a single copy alone is considered. $H_c$ generically has at most $\mO(1)$ redundancies when $p<p_c$\cite{Liu24}. In this case, the free energy (which is proportional to $-\log\mZ_R$) is analytic as a function of $\alpha$, since $\mZ_R$ involves only a sum of $\mO(1)$ finite numbers. 

Physically, in the absence of redundancies between the parity checks, each of them can independently be satisfied, allowing $H$ to be treated as a free model. This can be seen as follows.

In a model of $N$ bits with $M\leq N$ parity checks, collected in a parity check matrix $H$, we define a set of $\tau_{i=1\dots M}$ variables 

\[\tau_i \equiv H_{ij} s_j.\]

Clearly, in terms of these $\tau$ variables,
\[ H = \sum\limits_i \tau_i,\] and
\[\mZ = \sum\limits_{\qty{\tau_i}} \exp(-\beta\sum\limits_i \tau_i),\]
excluding those $\vec{\tau}$ configurations for which no $\vec{s}$ of the real, physical bits solves $\vec{\tau} = H\vec{s}$.

For example, consider the Ising model $H^I$ on a two dimensional lattice. One can similarly define $\tau_{ij}\equiv s_i \bsum[] s_j$ on the links between spins $i$ and $j$. Not every $\tau$ configuration is possible, since the sum of $\tau$ around a plaquette has to be 0; if $\tau_{1,2,3,4}$ denote the 4 parity checks around a plaquette, the configuration
\[\vec{\tau} = \mqty(1&0&0&0)^T \]
is unphysical. 
The $\tau$ bits are unconstrained -- and can independently be summed over -- only if the $\tau_i$ are linearly-independent; that is,
\[
\begin{aligned}
\bsum[i] a_i\tau_i =0 \iff a_i = 0\\
\implies a_iH_{ij}s_j =0 \iff a_i = 0\\
\implies \vec{a}^TH = 0 \iff \vec{a} = 0.
\end{aligned}\]

Alternatively, the image of $H$ can contain every vector in $\mF^M$ if and only if the rank of $H$ is $M$. In this case, the rows of $H$ are linearly independent and thus, $H$ has no constraints. Moreover, by the rank-nullity theorem, there are $k=N-M$ independent configurations which result in $\vec{\tau}=0$. From the linearity of $H$, we further conclude that each configuration $\vec{\tau}$ has the same multiplicity -- there are $2^k$ physical $\vec{s}$ for each $\vec{\tau}$. 

Thus, since $H_c$ has $\mO(1)$ redundancies, all but a vanishing fraction of the $\tau_i$ can be independently summed over. The free energy then simplies to 
\[
\begin{aligned}
\frac{1}{N}\log\mZ^{(1)} &= \frac{1}{N}\qty(\log 2^k\qty(\sum\limits_\tau \exp(-\beta \tau))^M + \dots)\\
&=k + \frac{M}{N} \log 2\cosh(\beta) + \frac{1}{N}\qty(\dots).
\end{aligned}
\]

$\dots$ refers to the term that stems from the redundancies. However, since it is accompanied by a $\frac{1}{N}$ pre-factor, this bounded sum of $\mO(1)$ terms vanishes in the thermodynamic limit, and the system undergoes no finite-temperature phase transition.

\subsection{Application to the Finite-Temperature Transition}

We now detail the implementation of the Kramers-Wannier transformation to the evaluation of $\mZ^{(2)}$. As a reminder, the set-up considered here involves two copies of a given realization of $H_c$ that share the same boundary. The parity check matrix, denoted $H_c^{(2)}$, can be written in the following fashion

\begin{equation}
\begingroup%
    \def\svgwidth{0.5\linewidth}
  \makeatletter%
  \providecommand\color[2][]{%
    \errmessage{(Inkscape) Color is used for the text in Inkscape, but the package 'color.sty' is not loaded}%
    \renewcommand\color[2][]{}%
  }%
  \providecommand\transparent[1]{%
    \errmessage{(Inkscape) Transparency is used (non-zero) for the text in Inkscape, but the package 'transparent.sty' is not loaded}%
    \renewcommand\transparent[1]{}%
  }%
  \providecommand\rotatebox[2]{#2}%
  \newcommand*\fsize{\dimexpr\f@size pt\relax}%
  \newcommand*\lineheight[1]{\fontsize{\fsize}{#1\fsize}\selectfont}%
  \ifx\svgwidth\undefined%
    \setlength{\unitlength}{452.99827275bp}%
    \ifx\svgscale\undefined%
      \relax%
    \else%
      \setlength{\unitlength}{\unitlength * \real{\svgscale}}%
    \fi%
  \else%
    \setlength{\unitlength}{\svgwidth}%
  \fi%
  \global\let\svgwidth\undefined%
  \global\let\svgscale\undefined%
  \makeatother%
  \begin{picture}(1,0.65238907)%
    \lineheight{1}%
    \setlength\tabcolsep{0pt}%
    \put(0,0){\includegraphics[width=\unitlength,page=1]{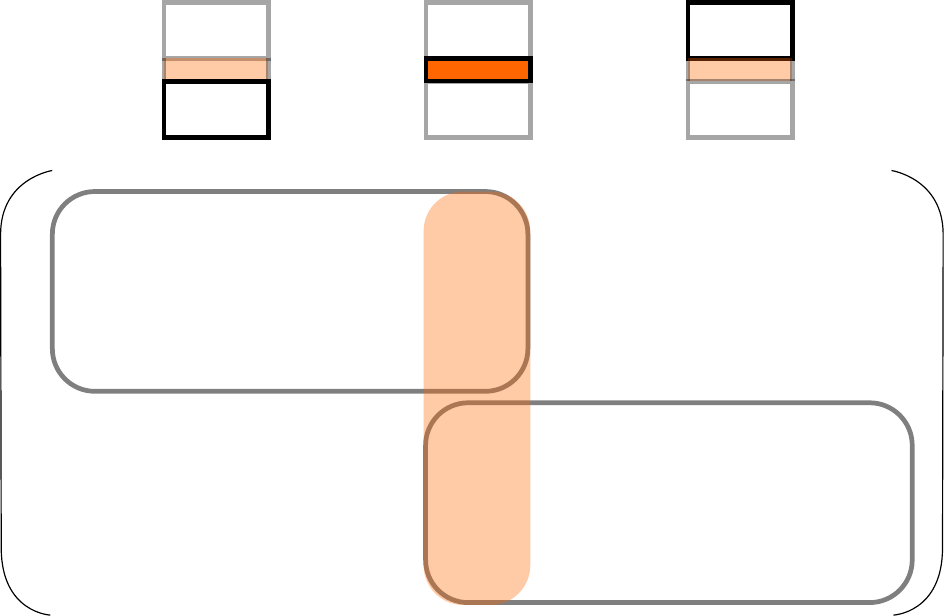}}%
    \put(0.22597775,0.32){\color[rgb]{0,0,0}\makebox(0,0)[lt]{\lineheight{1.25}\smash{\begin{tabular}[t]{l}{\LARGE $H_c$}\end{tabular}}}}%
    \put(0.69571963,0.1){\color[rgb]{0,0,0}\makebox(0,0)[lt]{\lineheight{1.25}\smash{\begin{tabular}[t]{l}{\LARGE $H_c$}\end{tabular}}}}%
    \put(0.71,0.32){\color[rgb]{0,0,0}\makebox(0,0)[lt]{\lineheight{1.25}\smash{\begin{tabular}[t]{l}{\LARGE $0$}\end{tabular}}}}%
    \put(0.22597775,0.1){\color[rgb]{0,0,0}\makebox(0,0)[lt]{\lineheight{1.25}\smash{\begin{tabular}[t]{l}{\LARGE $0$}\end{tabular}}}}%
  \end{picture}%
\endgroup%
.
    \label{eq:H2c}
\end{equation}

The columns are indexed by the bits in the bulk of each copy (left- and right-most columns), and the shared boundary bits (orange columns). The bulk bits participate in their respective parity checks, while the boundary bits participate in parity checks arising from both copies of $H_c$. To recap our notation, we have

\begin{equation}
    \centering
    \begin{tabular}{|c|c|}
    \hline
    \textbf{Variable} & \textbf{Number of} \\
    \hline
    $N$  &  Qubits in one copy\\
    $M$  &  Parity checks in one copy\\
    $Q$ & Redundancies (when BCs are imposed)\\
    \hline
    \end{tabular}.
\end{equation}

The parity checks in a single copy of $H_c$ are centered on all bits except for those at $t=1$ and $t=T$, giving us $N-2L$ parity checks. However, the $L$ bits at $t=1$ have a definite initial value (set by the initial conditions), For the system described by $H_c^{(2)}$, we have $M^{(2)} = 2M$ parity checks in total. The 2 copies share the bits at the boundary, tentatively resulting in $2N-2L$ bits. Additionally, we choose initial conditions corresponding to $s_{x,t=1} = 0$ for each copy; these bits can be removed from the system, giving us $2N-4L = 2M$ bits in total. Thus, $H_c^{(2)}$ is a $2M\times2M$ binary matrix, with a structure as in \cref{eq:H2c}.

\begin{figure}
    \centering
    \includegraphics[width=0.75\linewidth]{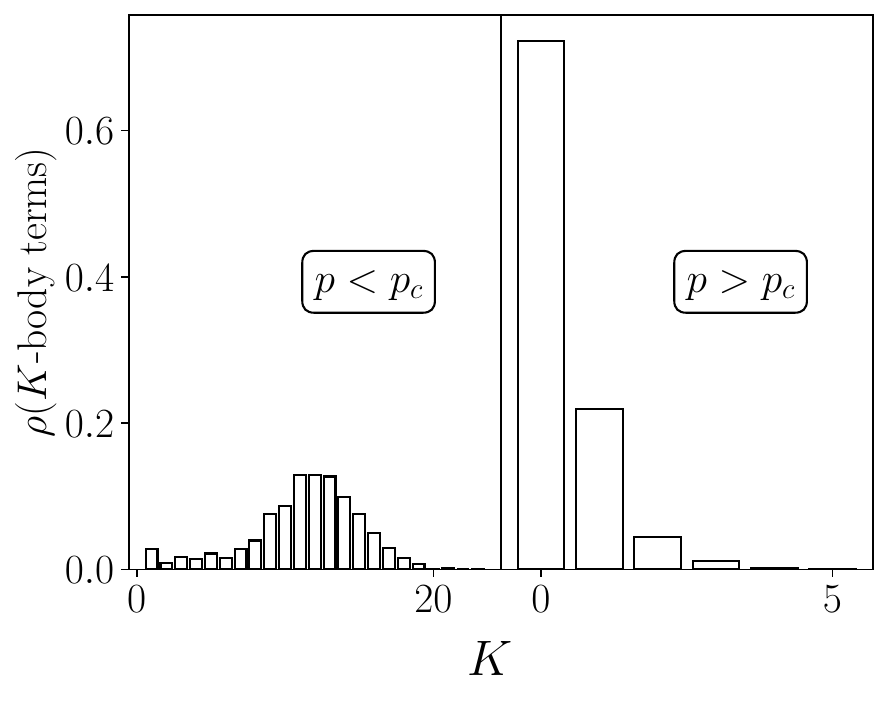}
    \caption{The distribution of the number of bits that are involved in each parity check of $R^T$, as we tune $p$. When $p>p_c$, most $q$-bits do not participate in \textit{any} parity check, while those that do, are involved in few-body interactions that generally cannot host a finite-temperature phase transition. When $p<p_c$, however, the interactions most commonly involve $\mO(L)$ spins. Furthermore, these interactions are frustrated -- there are many more parity checks than spins. Data are shown for a system with linear size $L=40$. }
    \label{fig:k-dist}
\end{figure}

For a given realization, we numerically obtain $\ker_L H_c^{(2)}$, whose basis vectors can be collected in a $Q\times2M$ matrix $R_{(2)}$ such that $R_{(2)}H_c^{(2)} = 0$. This greatly simplifies our problem, since $Q\sim \mO (\sqrt{N})$. Moreover, $R_{(2)}^T$ is the parity check matrix of a (nearly) all-to-all coupled model of these $Q$ spins when $p<p_c$. The change in the structure of $R_{(2)}$ as we tune $p$ is shown in \cref{fig:k-dist}.

At zero temperature, the expression for entanglement entropy took the form of a difference between the number of $\sX$ in a replicated space. We were however able to rewrite that expression in terms of the $\sX$ of a \textit{single} copy, which was more physically illuminating in terms of the local extent of $\sX$ on $A$ and $\overline{A}$.

We now describe a similar interpretation of the redundancies $\sR$, which, like the $\sX$, form a group, labelled $R$. For a single copy of the system, we define $\sR$ to be the solutions to the equation
\[\sR H|_{t<T-1} = 0, \]
where $H|_{t<T-1}$ denotes the parity check matrix $H$ \textit{without} the columns corresponding to the bits at the boundaries. For every $\sR$, $\mqty(\mrule&\sR&\mrule\sR\mrule)$ is a redundancy of $H_c^{(2)}$. This can be seen as follows:

$\sR H$ results in a vector which is zero in the bulk; this identically holds for both copies of $H_c$ in \cref{eq:H2c}. $(\sR H)|_\del$ is identical for both copies as well. Since the boundaries are shared between copies 1 and 2, the boundary configurations are added to each other, resulting in \[\mqty(\mrule&\sR&\mrule\sR\mrule) H_c^{(2)} = 0\].

As with the logical operators, the subgroup of $R$ which satisfies $\sR H = 0$ constitute the ``bulk" redundancies (i.e.~the redundancies between the parity checks of a single copy of $H$). We term this group $R_B$. Additionally, we can further define the quotient group 
\[R_\del = R/R_B,\]
which we analogously term the ``boundary" redundancies -- these are linear combinations of parity checks which act only on boundary bits.

\begin{figure}
    \centering
    \includegraphics[width=0.75\linewidth]{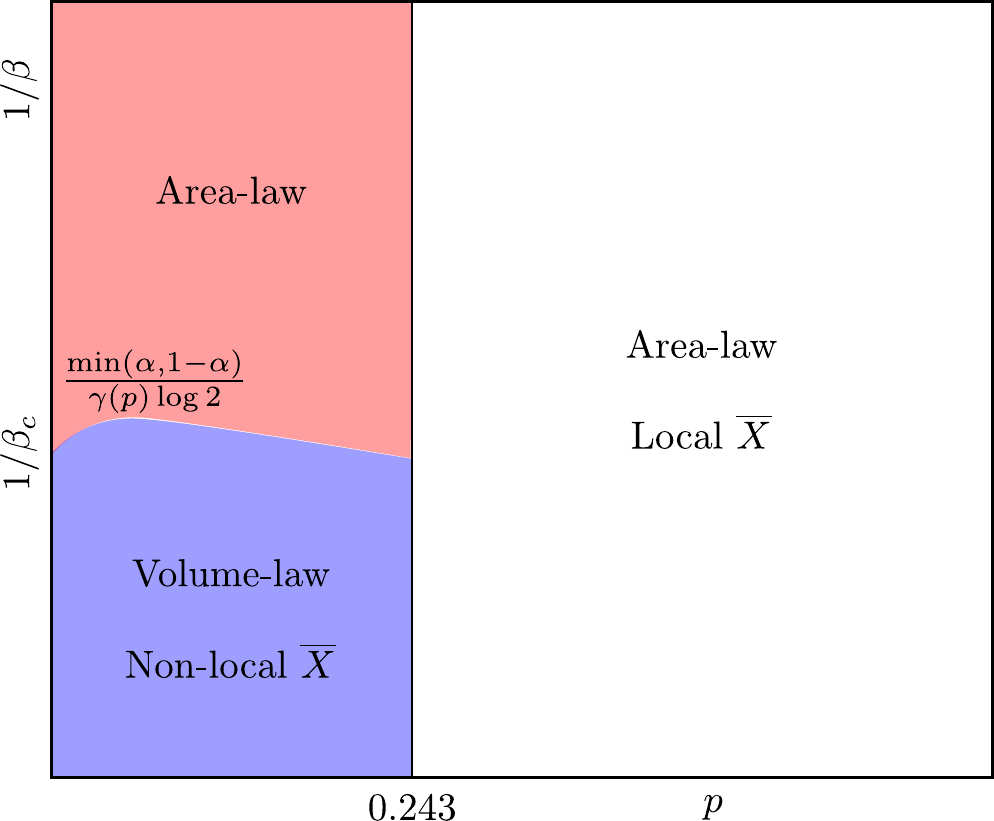}
    \caption{The complete phase diagram as both $\beta$ and $p$ are varied. When $p<p_c$, the system appears to undergo a volume-to-area law phase transition as the temperature is increased. This finite-temperature phase transition stems from the non-locality of the interactions in the dual model, phrased in terms of redundancies of $H_c^{(2/4)}$. When $p>p_c$, however, the system remains in a trivial, area-law phase, with redundancies involving only $\mO(1)$ bits, each; such a system cannot generally undergo a finite-temperature phase transition. }
    \label{fig:2dphaseD}
\end{figure}

To relate this back to the entanglement entropy under weak measurements, one needs to calculate $R_{(4)}$, which offers a similarly simplified expression for $\mZ^{(4)}$. $\se$ is given by a difference of free energies $\Delta F$.
\[\Delta F = 2\log\mZ^{(2)}-\log\mZ^{(4)}.\]
The complete phase diagram of our model, as functions of both the inverse temperature/measurement strength $\beta$ and the doping/measurement rate $p$, is described in \cref{fig:2dphaseD}. 

In conclusion, we have shown in this section that a thermal phase transition is \textbf{forbidden} in a single copy of the model; while a phase transition can occur in this replicated model. A compelling free energy argument was presented in Sec.~\ref{sec:free_energy}, suggesting that the boundary excitations can eventually proliferate and destabilize the ordered phase, albeit only below a critical $\beta_c$. We identified that the linear in $L$ scaling of the both the entropy and the energy cost is responsible for the competition that leads to this phase transition.

It remains to be seen what the exact structure of the boundary redundancies is, and how this relates to the putative thermal phase transition. A deeper exploration of how these redundancies arise and a full treatment of the model at finite $\beta$ is left to future work.

\section{Connection to a Classical Spin Model}
We stated in the main text that the quantum model in our work is directly related to a previously studied plaquette model of Ising spins. This classical Hamiltonian $H$ consists of $q$-spin terms placed on each site of the square lattice with probability $p (q=1)$ or $1-p (q=5)$.

Heuristically, the $q=5$ term can be understood as follows: At each time step, CNOT gates act on a shared target bit that stores the combined parity of all the control bits. Consider a target bit $a$ in a state $s_a$, and a set of control bits $b_{-1}, b_0, b_1$. After the gates $\prod_{j} {\rm CNOT}_{b_j,a}$ act on $a$, its state is  $s_c = s_a + \sum\limits_j s_{b_j} \mod 2$. Thus, the final state $s_c$ satisfies the constraint $s_a + \sum\limits_j s_{b_j} + s_c=0 \mod 2$. By identifying the sites $(x,t-1)\equiv a, (x,t+1)\equiv c$ and $(x+j-1,t)\equiv b$, we obtain the 5-bit term of $H_c$ at each site,
\begin{equation}
	s_{x,t+1} + \sum\limits_j s_{x+j,t} + s_{x,t-1}=0 \mod 2,
	\label{eq:5bodyCons}
\end{equation}
with $j=\pm1, 0$. The $q=1$ single-body term corresponds to measurements, since both measurements and the $q=1$ term involve fixing the state of the measured qubit, independent of its neighbours. We now present two methods -- one dynamical, and one based on tensor networks -- that explore this mapping in more detail. In particular, the dynamical mapping will show that the measurement outcome does not alter the classical mapping.

\subsection{Cellular Automaton}

The first method follows the evolution of each of the stabilizers describing the state $\ket{\psi}$. For more details about the stabilizer formalism, including expressions for the entanglement entropy, we refer readers to Refs.~\cite{Aaronson_2004, MIPT1}.

We begin by considering a given stabilizer whose elements are either $Z$ or $I$. This stabilizer can equivalently be described by a binary vector $S(t)\in\mathbb{F}_2^{2L}$ whose elements are labelled $\qty{s_{x,a,t},s_{x,b,t}}$, with the understanding that the stabilizer acts on site $(x,a)$ as $Z_{(x,a)}^{s_{x,a,t}}$ at time $t$.

CNOT gates act in the following fashion on a given stabilizer -- if $c$ and $t$ denote the coordinates of the control and the target, respectively, we have

\begin{equation}
    \begin{aligned}
        Z_c&\to Z_c,\\
        Z_t&\to Z_t Z_c,
    \end{aligned}
\end{equation}

or graphically,
\[
\includegraphics[width=0.25\textwidth]{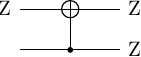}.
\]

In this example, the two $Z$ stabilizers can equivalently be represented by a vector of Boolean variables --
\begin{equation}
    \begin{aligned}
        Z_c \equiv (s_c=1,s_t=0),\\
        Z_t \equiv (s_c=0, s_t=1).
    \end{aligned}
\end{equation}

The action of the CNOT gate results in the following update rule, where for each Boolean vector $\vec{s}$, we have
\begin{equation}
    s_c \to s_t + s_c \mod 2.
\end{equation}

Returning to the quantum circuit, the control of each CNOT gate lies on the $b$ qubits, and the target on the $a$ qubits. Thus, the update rule for one timestep at each site $x$ is

\begin{equation}
    \begin{aligned}
        &s_{x,b,t+1} = s_{x,b,t} + s_{x,a,t} + s_{x-1,a,t} + s_{x+1,a,t}\\
    \implies &s_{x,b,t+1} + s_{x,b,t} + s_{x,a,t} + s_{x-1,a,t} + s_{x+1,a,t} = 0;\\
    &s_{x,a,t+1} = s_{x,a,t}.
    \end{aligned}
\end{equation}

Incorporating the effect of the SWAP gate at time $t+1$ -- i.e.~after the CNOT gates -- we have
\begin{equation}
\begin{aligned}
    s_{x,a,t+1} &= s_{x,b,t} + s_{x,a,t} + s_{x-1,a,t} + s_{x+1,a,t};\\
    s_{x,b,t+1} &= s_{x,a,t}.
\end{aligned}
\label{eq:P5term}
\end{equation}

Measurements, on the other hand, lead to the constraint that $s_{x,b,t} = 0$, since an X measurement on a site simply requires that the stabilizer $S$ must have no $Z$ weight on that site. We ignore the effect of the accompanying Z measurement for the moment.

Since the value of $s$ on the $a$ sublattice remains fixed between applications of CNOT gates, we can discard the sublattice indices in \cref{eq:P5term} using the prescription (with the $x$ label suppressed)

\[
\begin{aligned}
    (a,t) \equiv t;\\
    (b,t) \equiv t-1,
\end{aligned}
\]

so that \cref{eq:P5term} returns the value of $s_{x,t+1}$ as a function of the values at $t, t-1$. Note that although $(b,t+1)$ and $(a,t)$ refer to the same site, there is no ambiguity since at each time step, $s_{x,b,t+1} \equiv s_{x,a,t}$. Finally, the update rules \cref{eq:P5term} can be written in terms of $L\times T$ (instead of $2L\times T$) variables $\qty{s_{x,t}}$ to give

\[s_{x,t+1} = s_{x,t-1} + s_{x,t} + s_{x-1,t} + s_{x+1,t}.\] 

These reinterpreted couplings can now be cast into a family of 2d classical Hamiltonians $H_c$, whose ground states are those which satisfy all these 5- and 1-body constraints. To complete the definition, let $q_{x,t}\in\qty{1, 5}$ be a random variable defined on each site, where $q=1 (5)$ indicates the absence (presence) of the 5-body interaction centered at site $(x,t)$; $q=1 (5)$ equivalently indicates the presence (absence) of a measurement in the corresponding quantum dynamics. The Hamiltonian for a particular realization of $\qty{q}$ is

\begin{equation}
    H_c\qty(\qty{q}) = \sum\limits_{x,t} \delta_{q_{x,t},1} \sigma_{x,t} + \delta_{q_{x,t},5} \sigma^5_{x,t},
    \label{eq:Hc}
\end{equation}

where $\sigma^5_{x,t} \equiv \sigma_{x,t}\sigma_{x+1,t}\sigma_{x-1,t}\sigma_{x,t+1}\sigma_{x,t-1}$. Treating $H_c$ as the parity check matrix of a classical code, the ground states form codewords, while the $s_{x,t}$ obtained through the automaton process determine the spins that need to be flipped to obtain (a subset of) the codewords, starting from the $\sigma=1$ state. 

The role of the $Z$ measurement becomes evident when considering the automaton process. A $Z$ measurement that anti-commutes with any $X$ stabilizer results in the destruction of that $X$ stabilizer, as well as the creation of a \textit{new} single-site $Z$ stabilizer. In the classical picture, the effect of the Z measurement accompanying the X measurement is to create a new ``symmetry operator\footnote{Let $\qty{s}$ denote a collection of sites whose spins need to be flipped in order to reach a non-trivial codeword from the all-up state. The corresponding symmetry operator is given by $\prod\limits_\qty{s} X$.}". A detailed exploration of the automaton process and its implications on the zero temperature partition function of $H$ were recently studied in \cite{Liu24}. The behavior of these operators bear intriguing similarities to that of entanglement in certain settings.

\subsubsection{Slow Dynamics}

This connection explains the slow dynamics of the model as well. Recall that we could write the state emerging from the quantum circuit as

\begin{equation}
    \ket{\psi\qty(T)} \propto \sum\limits_{s_T}\mZ\qty(s_T)\ket{s_T}.
\end{equation}

Clifford states are equal weight superpositions over the computational basis. Thus $\mZ(s_T)$ is independent of $s_T$, provided that $\mZ(s_T)\neq0$; that is, the boundary conditions $s_T$ should admit at least one valid ground state of $H_c$. $PE_Z(\ket{\psi}$ then is the logarithm of the total number of such valid boundary conditions. Obtaining this number for $H_c$ is related to its symmetry group $G$.

$G_{\rm bulk}$, the group of all symmetry operators that act purely in the bulk (away from rows $T-1$ and $T$), forms a subgroup of $G$. The quotient group $G_\partial \equiv G/G_{\rm bulk}$, on the other hand, contains the operators that do alter the boundary (modulo bulk operators), so we term $G_\partial$ the boundary symmetry group. If $s$ is a classical ground state, then so is $X[s]$, $\forall X \in G_\partial$, but with \textit{different} boundary conditions. Thus, the number of distinct boundary conditions that lead to a ground state of $H_c$ is the size of $G_{\partial}$. As was the case with $G$, one can find $N_X=\log |G_\partial|$ independent \textit{boundary} symmetry operators $X_j$ that generate $G_\partial$, so $PE_Z \equiv N_X$. We now study how $N_X$ evolves with $T$, by studying the evolution of these boundary symmetry generators (``bSG").

Recall that there are no terms in $H_c$ that are centered on row $T$. The lattice is now extended to row $T+1$ by introducing couplings at row $T$. At $(x,T)$, if there is a 5-body term, then the configuration of a bSG at $(x,T+1)$ is completely determined by its configuration at rows $T$ and $T-1$. By contrast, a 1-body term constrains $X_{(x,T)} = \mathbb{I}$ for all the bSG. If there are $m$ bSGs such that $\qty(X_j)$ flips the spin at $(x,t)$ for $j=1\dots m$, we can redefine $X_j\to X_jX_1$ for $j\neq1$ and discard $X_1$. These new $X_j$ generate a subgroup of $G_\partial$ which is not supported on $(x,T)$. Additionally, $s_{x,T+1}$ can be 0 or 1 (since it does not feature in $H_c$), creating a new symmetry operator $\widetilde{X} = X_{x,T+1}$. Ultimately, the bSGs for $G_\partial$ are $\qty{\widetilde{X},X_j}$, with $j\geq2$ if an operator is discarded as above.

We readily see that $N_X$ increases by 1 at site $(x,T)$ only if there is a 1-body term at $(x,T)$ \textit{and} no other bSG is supported on that site. When $p<p_c$, a finite density of bSG are non-local with typical length $l=\Omega(L)$ at times $T\gtrsim L$. The probability that there is no other bSG at a site is $(1-\frac{l}{L})^{N_X}$, as in the main text. This exponentially small rate leads to a logarithmic growth of $N_X$ at late times, and thus, of $S_A$.

\subsection{Tensor Network Method}

An alternate prescription comes from considering the tensor network associated with the quantum circuit. We begin by introducing a ``green" tensor

\begin{center}
    \begin{tikzpicture}
    \foreach \t in {90,210,330}{
        \draw(0,0) -- +(\t:0.5cm) ;
    }
    \draw (0,0) node [circle,draw,fill=green!20]{};
    
    \node at (1,0) {$\equiv$};
    
    \begin{scope}[xshift=2cm]
        \foreach \t in {90,210,330}{
            \node at (\t:0.35cm) {$\ket{0}$};
        }
    \end{scope}
    
    \node at (3,0) {$+$};
    
    \begin{scope}[xshift=4cm]
        \foreach \t in {90,210,330}{
            \node at (\t:0.35cm) {$\ket{1}$};
        }
    \end{scope}
\end{tikzpicture}
\end{center}

which represents the GHZ state on its legs. This definition can be extended to include any number of legs, and the contraction of any two green tensors results in a single green tensor with a larger GHZ state on its uncontracted legs. Diagramatically,

\begin{center}
\begin{tikzpicture}
    \draw (0,-0.5) -- +(0,1);
    \draw (0.5,-0.5) -- +(0,1);
    \draw (0,0) node[circle,draw,fill=green!20] {} -- (0.5,0) node[circle,draw,fill=green!20] {};
    
    \node at (1.25,0) {=};
    
    \begin{scope}[xshift=2cm]
        \foreach \t in {45,135,225,315}{
            \draw (0,0) -- (\t:0.5cm);
        }
        \node[circle,draw,fill=green!20] at (0,0){};
    \end{scope}
\end{tikzpicture}
\end{center}

A ``red" tensor is one obtained by applying a Hadamard gate on each leg of a green tensor.

\begin{center}
\begin{tikzpicture}
    
    \foreach \t in {90,210,330}{
        \draw(0,0) -- node [rectangle,draw,fill=yellow,inner sep = 2pt,pos=0.65,rotate=\t]{} +(\t:0.75cm) ;
    }
    \draw (0,0) node [circle,draw,fill=green!20]{};
    
    \node at (1,0.25) {=};
    
    \begin{scope}[xshift=2cm]
        \foreach \t in {90,210,330}{
            \draw(0,0) -- +(\t:0.5cm) ;
        }
        \draw (0,0) node [circle,draw,fill=red!20]{};
    \end{scope}

\end{tikzpicture}
\end{center}

A fusion of red tensors results in another red tensor, analogously. In the computational basis, the green tensor constrains all its indices to have the same value (either 0 or 1), while the red tensor allows only an even number of its indices to take the value 1. By defining (classical) spin variables $s \equiv (-1)^n$ from the index variables $n \in \qty{0,1}$, the red tensor can be reinterpreted as an Ising-like coupling between the spins on its legs

\begin{equation}
\begin{tikzpicture}
    \foreach \t/\x in {90/1,210/2,330/3}{
        \draw(0,0) -- ++(\t:0.5cm) node[pos=1.5] {$s_{\x}$} ;
    }
    \draw (0,0) node [circle,draw,fill=red!20]{};
    
    \node at (1.5,0) {$\equiv$};
    \node[font=\huge] at (4,0) {$\frac{1 + s_1s_2s_3}{2}$};
\end{tikzpicture}.
\label{eq:RTens}
\end{equation}

With these tensors, one can write a CNOT gate as

\begin{center}
\begin{tikzpicture}
    \draw (0,-0.5) -- +(0,1);
    \draw (1,-0.5) -- +(0,1);
    \draw (0,0) node[circle,draw,fill=green!20] {} -- (1,0) node[circle,draw,fill=red!20] {};
\end{tikzpicture},
\end{center}
where the green tensor corresponds to the control. The tensor network for a time-step without measurements, with the $b$ sites ahead of $a$, is

\begin{center}
\begin{tikzpicture}
    
    \foreach \x in {0,1,2}{
        \draw (\x,0,0) node[circle,draw,fill=red!20] (R\x) {} -- (\x,0,1) node[circle,draw,fill=green!20] (G\x){};
        \draw (R\x.north) -- +(0,0.5);
        \draw (R\x.south) -- +(0,-0.5);
        
        \draw (G\x.north) -- +(0,0.5);
        \draw (G\x.south) -- +(0,-0.5);
    }
    
    \draw (R1.east) -- (G2.west);
    \draw (R0.east) -- (G1.west);
    
    \draw (R1.west) -- (G0.east);
    \draw (R2.west) -- (G1.east);
    
    \foreach \x in {0,1,2}{
        \draw (\x,0.5,0) node[rectangle,inner sep=0pt,fill=white] {\tiny$\boldsymbol{\times}$} -- +(0,0,1) node[rectangle,inner sep=0pt,fill=white] {\tiny$\boldsymbol{\times}$};
    }

\end{tikzpicture}.
\end{center}

Each pair of green and red tensor is recast as a doubled tensor, with the even (odd) legs corresponding to those of the green (red) tensor, as counted clockwise from above. The red tensor ensures that the parity of all the spins in the five body term is positive. The green tensor replicates each index five times, since each spin participates in at most five checks -- one from each of its neighbours, and one for itself. With the even (odd) legs colored green (red), we have the following double tensor

\begin{equation}
    \includegraphics[width=0.5\textwidth]{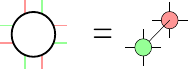},
\end{equation}
whose elements are given by

\begin{center}
\begin{tikzpicture}

    \draw (-0.75,0.2) -- +(1.5,0) node[anchor=mid west] {$s_{x+1,t}$};
    \draw (-0.75,-0.2) node[anchor=mid east] {$s_{x-1,t}$} -- +(1.5,0);
    
    \draw (0.2,-0.75) node[anchor=north west,inner xsep=-1pt] {$s_{x,t-1}$} -- +(0,1.5);
    \draw (-0.2,-0.75) -- +(0,1.5) node[anchor=south east,inner xsep=-8pt] {$s_{x,t+1}$};
    
    \draw[fill=white] (0,0) circle [radius=1.2em];

    \node[anchor=mid west] at (2cm,0) {$=\delta_{(s_{x+1,t}s_{x,t-1}s_{x-1,t}s_{x,t+1}s_{x,t}),1}$};
\end{tikzpicture}
\end{center}

The unlabelled indices all correspond to the spin value $s_{x,t}$. This tensor is placed on the site $(x,t)$ on a two dimensional square lattice. The somewhat peculiar labelling of the indices makes clear how neighboring tensors are to be contracted, i.e. for two tensors located at $(x,t)$ and $(x'\equiv x+1,t)$, the index of the first tensor labelled $s_{x+1,t}$ has to join the index $s_{x',t}$ of the second tensor. For two neighboring tensors (independent of their relative angles), the contraction proceeds as

\begin{center}
\begin{tikzpicture}

    \draw[rounded corners] (-0.75,-0.2) -- +(1.5,0) -- (1,0) -- (1.25,0.2) -- +(1.5,0);
    \draw[rounded corners] (-0.75,0.2) -- +(1.5,0) -- (1,0) -- (1.25,-0.2) -- +(1.5,0);
    
    \draw (0.2,-0.75) -- +(0,1.5);
    \draw (-0.2,-0.75) -- +(0,1.5);
    
    \draw[fill=white] (0,0) circle [radius=1.2em];

    \begin{scope}[xshift=2cm]
    
    \draw (0.2,-0.75) -- +(0,1.5);
    \draw (-0.2,-0.75) -- +(0,1.5);
    
    \draw[fill=white] (0,0) circle [radius=1.2em];
    \end{scope}
\end{tikzpicture}.
\end{center}

This is in keeping with the observation from the quantum circuit that red tensors are only connected to green tensors (and vice versa).

In our protocol, measurements are performed in the Z-basis on $b$ sites (green tensors) and the X-basis on $a$ sites. When the outcome of a Z measurement is $\ket{n}$, the other legs on the green tensor -- the even legs of the doubled tensor -- are constrained to be $n$. The effect on the red tensor is analogous, except now in the X-basis. Measurements thus disconnect the tensors on the measured sites from the rest of the network (and each other). The graphical representation of a tensor at site $(x,t)$, corresponding to Z and X measurements (with respective outcomes $n\in\qty{0,1}$ and $m\in\qty{+,-}$), is

\begin{center}
\begin{tikzpicture}
    
    \draw (-0.75,0.2) -- +(1.5,0) node[anchor=mid west] {$s_{x+1,t}$};
    \draw (-0.75,-0.2) node[anchor=mid east] {$s_{x-1,t}$} -- +(1.5,0);
    
    \draw (0.2,-0.75) node[anchor=north west,inner xsep=-1pt] {$s_{x,t-1}$} -- +(0,1.5);
    \draw (-0.2,-0.75) -- +(0,1.5) node[anchor=south east,inner xsep=-8pt] {$s_{x,t+1}$};
    
    \draw[fill=white] (0,0) circle [radius=1.2em];
    
    \node at (2.5,0) {$\equiv$};
    
    \begin{scope}[xshift=4cm]
        \node at (-0.75,0.2) {$\ket{n}$};
        \node at ($(-0.75,0.2)+(1.5,0)$) {$\bra{m}$};
        \node at (-0.75,-0.2) {$\ket{m}$};
        \node at ($(-0.75,-0.2)+(1.5,0)$) {$\bra{n}$};
        
        \begin{scope}[rotate=-90]
            \node at (-0.75,0.23) {$\ket{n}$};
            \node at ($(-0.75,0.23)+(1.5,0)$) {$\bra{m}$};
            \node at (-0.75,-0.23) {$\ket{m}$};
            \node at ($(-0.75,-0.23)+(1.5,0)$) {$\bra{n}$};
        \end{scope}
    \end{scope}
\end{tikzpicture}
\end{center}

Putting these tensors together, we arrive at a tensor network which exactly encodes the zero temperature partition function of the classical spin Hamiltonian defined in \cref{eq:Hc}. An immediate advantage of the tensor network formalism is that it remains valid even when the measurements are no longer projective.

\subsubsection{Extension to Finite-Temperature}

We begin by introducing weak (forced) measurements in the $X$-direction with a strength $\gamma$. This measurement now takes place at \textit{all} $b$ sites following the application of the CNOT gates. In the tensor network, this amounts to contracting each red tensor with the operator $\exp(\gamma X)$. As an example, we study the effect of this multiplication on the tensor of \cref{eq:RTens}, which we call $T_{s,t,u}$, with the relabelling $(s_1, s_2, s_3)\to (s,t,u)$.

\[\exp(\gamma X) \propto 1 + \tanh(\gamma)X,\]

where $X$ flips the spin $s\to -s$. As a result, we have 
\[\begin{aligned}
    \exp(\gamma X_{s',s})T_{s,t,u} &\propto \frac{1 + s.t.u}{2}+ \tanh(\gamma) \frac{1 - s.t.u}{2}\\
    &= \frac{1+\tanh(\gamma)}{2} + \frac{1-\tanh(\gamma)}{2} (stu)\\
    &\equiv \exp(\beta s_1 s_2 s_3),
\end{aligned}\]
where the last line follows from the redefinition 
\[\tanh(\beta) = \frac{1-\tanh(\gamma)}{1+\tanh(\gamma)},\] which simplifies to $\exp(-2\gamma) = \tanh(\beta)$.

What we have found so far is that multiplying the tensor $T$ with $\exp(\gamma X)$ results in a tensor $T(\gamma)$ that \textit{relaxes} the Ising constraint of $T$, so that the product $stu$ is not forced to be 1. Note that $T$ corresponds to $T(\gamma=0)$. Generalizing this result to the entire doubled tensor, this alters the Boltzmann weight of the red tensor from $\delta_{(s_{x+1,t}s_{x,t-1}s_{x-1,t}s_{x,t+1}s_{x,t}),1}$ (which only holds at $\beta\to\infty$), to 
$\exp(\beta s_{x+1,t}s_{x,t-1}s_{x-1,t}s_{x,t+1}s_{x,t})$, which is exactly the Boltzmann weight of the 5-body term at \textit{finite temperature}.

The recipe to alter the Boltzmann weight of the 1-body term, situated at a point $(x,t)$, is analogous. In this case, $X$-measurements remain projective, while the $Z$-measurements are replaced by a weak measurement of the form $e^{\beta Z}$. Recall that $Z$ ($X$) measurements are made on the top legs of the green (red) tensors, i.e.~on the qubits corresponding to $s_{x,t} (s_{x,t+1})$. The $X$ measurement disconnects the red tensors, leaving the green tensors alone.
\[
\includegraphics[width=0.5\textwidth]{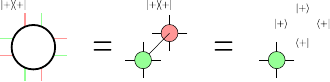}
\]

By performing the multiplication of $e^{\beta Z}$ with the green tensor, thereby contracting $e^{\beta Z}$ with $s_{x,t}$, the corresponding Boltzmann weight becomes $e^{\beta s_{x,t}}$
\begin{equation}
    \includegraphics[width=0.5\textwidth]{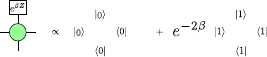}
\end{equation}

Thus, by introducing weak forced measurements in this specific fashion, we have established that the correspondence between the classical and the quantum model persists to finite temperatures. This provides an important first-step to a dynamical understanding of the finite-temperature phase transition in $H_c$ in the replica picture.

We note that the terminology used (specifically of red and green tensors and their fusions) is identical to that of ZX-calculus \cite{ZX1, ZX2}, a complete, graphical language that can be used to represent any quantum circuit (see \cite{ZX3} for a review).

\section{Additional Data}
\subsection{Measurement-induced Phase Transition}

\begin{figure}
    \centering
    \includegraphics[width=.7\linewidth]{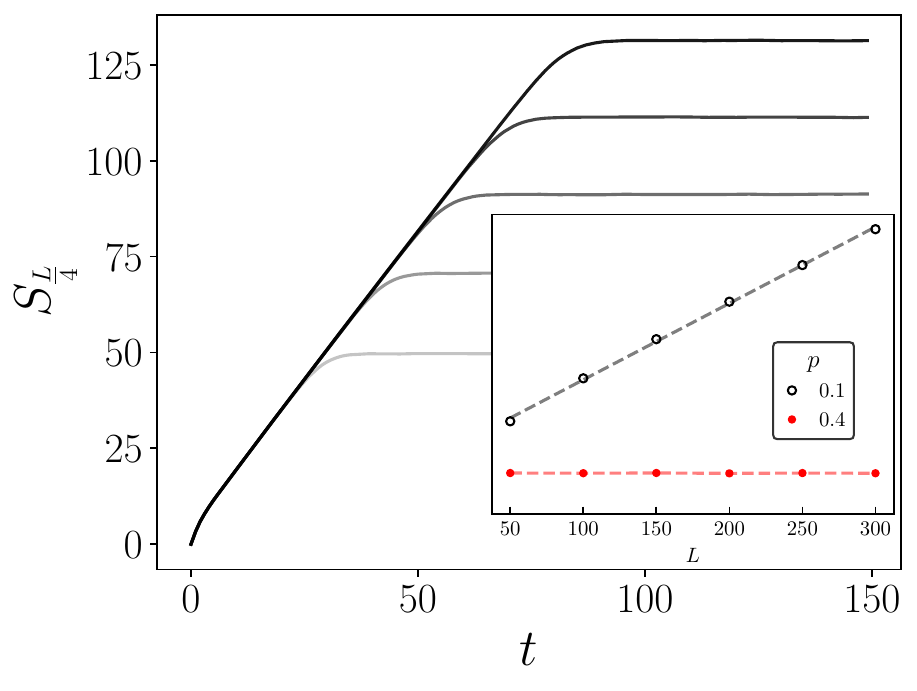}
    \caption{Growth of the entanglement entropy $S_{\frac{L}{4}}$ of a subsystem of size $L/4$, for system sizes $L\in\qty[100,300]$ (increasing in darkness from bottom to top), starting from a staggered initial state. $S$ saturates at an $L$ dependent value, indicating that the system is in a volume-law phase. (Inset) The steady state values of $S$ as a function of $L$ for $p<p_c$ (black) and $p>p_c$ (red). A linear fit yields $S_{\frac{L}{4}}\sim 0.23 L + 7.08$ in the volume law phase.}
    \label{fig:fast_dyn}
\end{figure}

The entanglement transition in the model, when starting from random initial conditions, is qualitatively similar to the volume- to area-law transition observed in generic unitary circuits interspersed with projective measurements. For completeness, we show in \cref{fig:fast_dyn} that the entanglement entropy grows linearly with time, finally saturating in an $L-$dependent value, for $p<p_c$. When $p>p_c$, the entanglement entropy is independent of system size, indicating that the system is in an area law.

\begin{figure}
     \centering
     \includegraphics[width=0.33\linewidth]{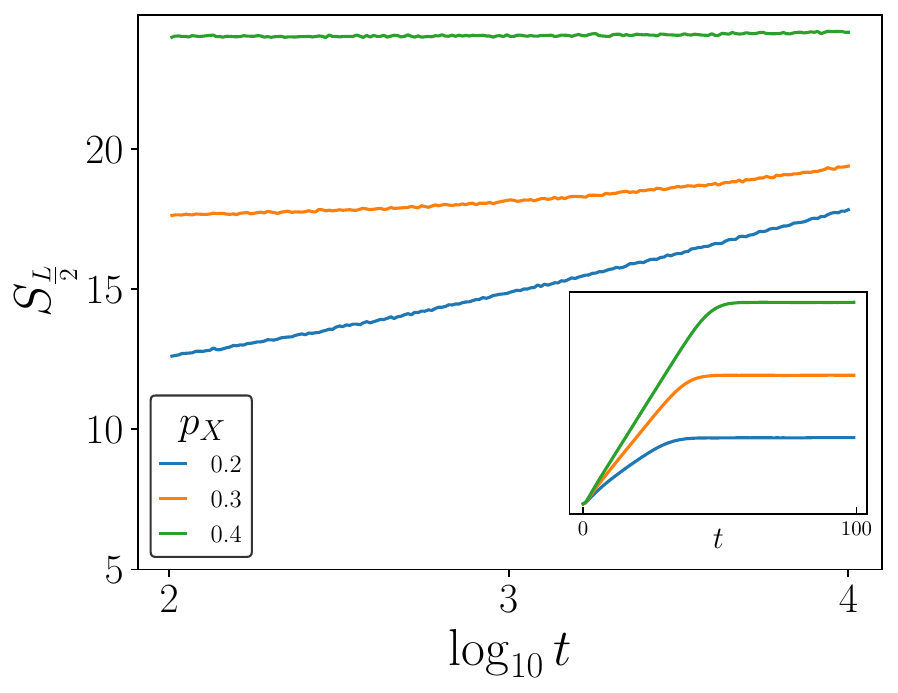}%
     \includegraphics[width=0.33\linewidth]{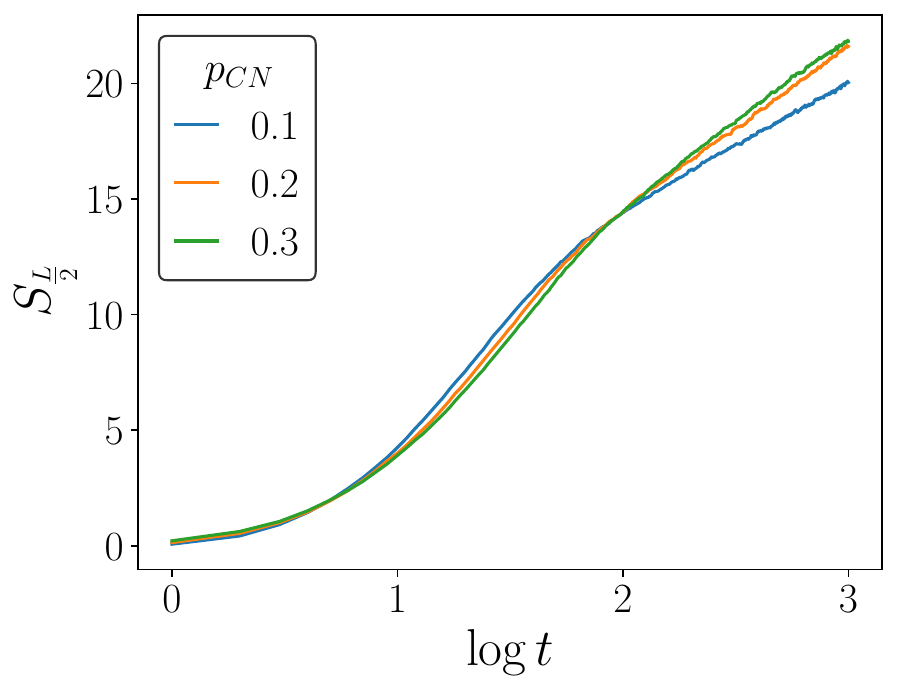}%
     \includegraphics[width=0.33\linewidth]{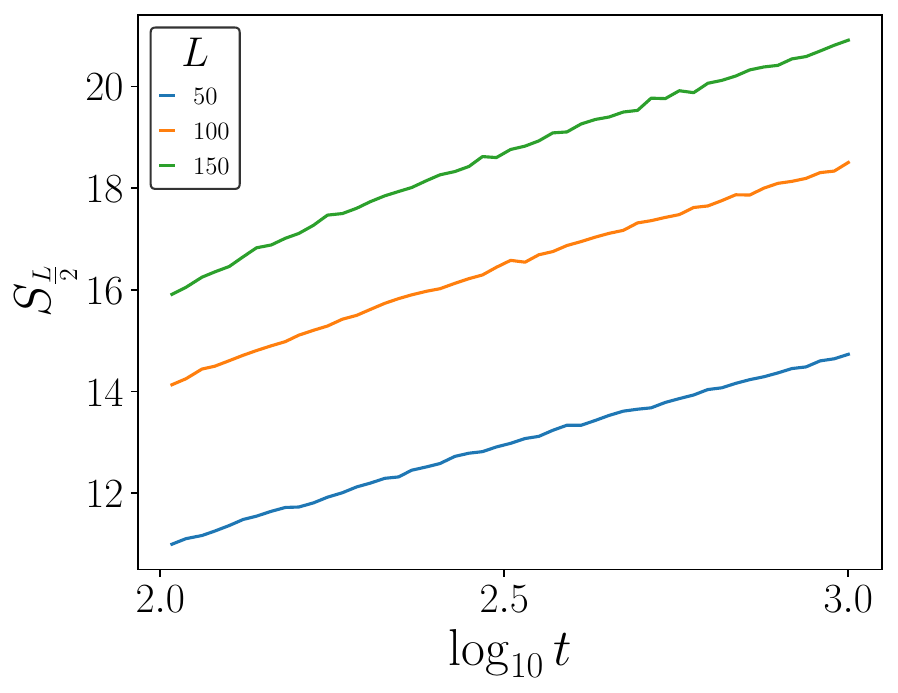}
     \caption{The dynamics remains robust even in the presence of different perturbations. (Left) The entanglement dynamics beginning from a random initial state, where $p_X$ of the qubits are initialized in the $X$ basis, while the rest are in the $Z$ basis. The entanglement growth is initially linear, and reaches a plateau for a duration proportional to $p_I L$. After this plateau, logarithmic growth sets in. (Center) Perturbing the $CNOT$ gates such that a fraction $p_{CN}$ of them are randomly chosen to be flipped (i.e.~the control is on the $b$ sublattice) does not hamper slow dynamics (Right) Logarithmic growth is observed even when a fraction (taken here to be 0.1) of the CNOTs are replaced by CZs.}
     \label{fig:perts}
 \end{figure}

\subsection{Slow dynamics across hybrid circuits}

The universality of slow dynamics extends beyond the specific model that we considered in the main text. In this section, we subject the model to various perturbations in the initial states, the measurement protocols, and in the unitary gates, while only requiring that the participation entropy conditions holds.

We begin by considering random initial conditions with unequal probabilities of $X$- and $Z$-polarized initial states; $p_X$ is the probability of placing an $X-$polarized initial state. We find that $S(t)\sim t$ at early times, quickly saturating to a $p_X$-dependent value that scales with the size of the system. However, this is not the true steady-state; merely a plateau in $S$. The entanglement remains at this plateau for an exponentially-long amount of time, after which \textit{logarithmic} growth sets in. The time at which logarithmic growth begins depends monotonically on $\min (p_X,1-p_X)$.

We also perturb the unitary gates by replacing a finite fraction of the $CNOT$ gates, randomly chosen at each time-step, with $CZ$ gates. Since we began from an all $Z$ state, $CZ$ does not alter the PE, ensuring logarithmic growth.

Finally, we randomly perturb the locations of the controls and targets of the $CNOT$ gates, such that controls (targets) no longer exclusively placed on the $a$ ($b$) sublattices. Again, logarithmic growth is persistent.

The participation entropy has been shown to be a useful indicator of entanglement, especially in automaton-like circuits where the action of the unitary gates on bitstrings in a specific basis is evident. While the PE has previously been used in circuits involving measurements and feedback to separate MIPTs from absorbing phase transitions \cite{VR_2023, VR_2025}, this work showcases its utility in a different setting.

\bibliography{reference}